\documentclass[12pt]{article}
\usepackage{amssymb}
\usepackage{amsmath}
\usepackage{cite}
\usepackage{axodraw}
\usepackage{hyperref}
\input xy
\xyoption{all}
\usepackage{enumitem}

\def\sqr#1#2{{\vcenter{\vbox{\hrule height.#2pt
            \hbox{\vrule width.#2pt height#1pt \kern#1pt
                  \vrule width.#2pt}\hrule height.#2pt}}}}

\def\sqra#1#2#3{{\vcenter{\vbox{\hrule height.#2pt
            \hbox{\vrule width.#2pt height#1pt \kern5pt 
#3
                  \vrule width.#2pt}\hrule height.#2pt}}}}

\numberwithin{equation}{section}

\numberwithin{table}{section}

\begin{document}

\begin{center}

{\large\bf Orbifolds by 2-groups and decomposition}

\vspace*{0.2in}

Tony Pantev$^1$, Daniel G. Robbins$^2$, Eric Sharpe$^3$,
Thomas Vandermeulen$^1$

\begin{tabular}{cc}
{\begin{tabular}{l}
$^1$ Department of Mathematics\\
David Rittenhouse Lab.\\
209 South 33rd Street\\
Philadelphia, PA  19104-6395 \end{tabular}}
&
{\begin{tabular}{l}
$^2$ Department of Physics\\
University at Albany\\
Albany, NY 12222 \end{tabular}}
\end{tabular} 

\begin{tabular}{c}
{\begin{tabular}{l}
$^3$ Department of Physics MC 0435\\
850 West Campus Drive\\
Virginia Tech\\
Blacksburg, VA  24061 \end{tabular}}
\end{tabular}

{\tt tpantev@math.upenn.edu},
{\tt dgrobbins@albany.edu},
{\tt ersharpe@vt.edu},
{\tt tvandermeulen@albany.edu}

\end{center}

In this paper we study three-dimensional orbifolds by 2-groups
with a trivially-acting one-form symmetry group $BK$.
These orbifolds have a global two-form symmetry, and so one expects
that they decompose
into (are equivalent to) a disjoint union of other three-dimensional theories,
which we demonstrate.
These theories can be interpreted as
sigma models on 2-gerbes, whose formal structures reflect properties
of the orbifold construction.

\begin{flushleft}
April 2022
\end{flushleft}

\newpage

\tableofcontents

\newpage

\section{Introduction}

Decomposition, the observation that a local quantum field theory is sometimes
a disjoint union of other local quantum field theories, has by now been
extensively studied since its initial observation
in \cite{Hellerman:2006zs} in two-dimensional gauge theories with trivially-acting
subgroups, see for example 
\cite{ajt1,ajt2,ajt3,t1,gt1,xt1,Caldararu:2007tc,Hellerman:2010fv,
Anderson:2013sia,Sharpe:2014tca,Sharpe:2019ddn,Tanizaki:2019rbk,Cherman:2020cvw,Cherman:2021nox,Robbins:2020msp,Robbins:2021ibx,
Robbins:2021lry,Robbins:2021xce,Eager:2020rra,Komargodski:2020mxz,Yu:2021zmu,
Nguyen:2021yld,Nguyen:2021naa,Honda:2021ovk,Huang:2021zvu}.
A few reviews can be found in
\cite{Sharpe:2006vd,Sharpe:2010zz,Sharpe:2010iv,Sharpe:2019yag,Sharpe:2022ene}.

Although decomposition was originally observed in two dimensional theories,
it has also been observed in four-dimensional theories, see
for example
\cite{Tanizaki:2019rbk,Cherman:2020cvw}.
The purpose of this paper is to
discuss examples in three dimensions, where it has not previously
been studied.

Globally, decomposition is expected to take place in any theory
in $d$ spacetime dimensions with a global $(d-1)$-form symmetry (possibly
realized noninvertibly) \cite{Tanizaki:2019rbk,Cherman:2020cvw}.  
One way to produce such a symmetry is via a suitable gauging.
In broad brushstrokes, gauging a trivially-acting $n$-form symmetry
results in a theory with a global $(n+1)$-form symmetry (distinct from
the quantum symmetry), so one can hope to produce a $d$-dimensional
theories with a decomposition by gauging a trivially-acting
$(d-2)$-form symmetry.

For example, ordinary
gauge theories with trivially-acting subgroups are
a source of examples in two dimensions, as mentioned above,
because such theories have a global
one-form symmetry (distinct from the quantum symmetry, and tied specifically
to the fact that the group acts trivially).  

In this paper, we study three-dimensional gauge theories with
gauged trivially-acting one-form symmetries.  Gauging the trivially-acting one-form symmetry leads to a global
two-form symmetry, hence, in three dimensions, a decomposition.

Specifically, in this paper we describe orbifolds of three-dimensional
effective\footnote{
We emphasize that because we often discuss orbifolds of three-dimensional
sigma models, we understand those sigma models as effective field
theories, not necessarily renormalizable theories.
Our methods also apply to more general three-dimensional theories,
such as, for example, Chern-Simons theories.
} field theories by 2-groups, which are extensions of ordinary
(here, finite) groups by
one-form symmetries.
(See for example \cite{baezlauda} for a mathematical introduction to 2-groups.
These structures have a long history in both math and physics,
see for example 
\cite{yetter,br,mackaay,porter,fmp,fhlt,Schommer-Pries:2011vyj,Baez:2005sn,Nikolaus:2011zg,Freed:1994ad,Baez:2010ya,Sati:2008eg,Pfeiffer:2003je,Frohlich:2009gb,Carqueville:2012dk,Carqueville:2013mpa,Brunner:2013ota,Brunner:2014lua,Carqueville:2015pra,Schreiber:2008uk,Sati:2009ic,Fiorenza:2010mh,sw,Kim:2019owc,Fiorenza:2012tb,Fiorenza:2013jz} 
for a few older instances,  
and 
\cite{Sharpe:2015mja,Cordova:2018cvg,Benini:2018reh,Cordova:2020tij,DelZotto:2020sop,Iqbal:2020lrt,Lee:2021crt,Fiorenza:2020iax,Sati:2020nob,Apruzzi:2021vcu,Bhardwaj:2021wif,Apruzzi:2021mlh,DelZotto:2022joo,Cvetic:2022imb} 
for some more recent physics descriptions and
applications of 2-groups.)
When those one-form symmetries act trivially, gauging them results in a global
two-form symmetry, hence a decomposition as above, which we will check
explicitly.

We begin in section~\ref{sect:ordinary-case} by reviewing two-dimensional
orbifolds by
central extensions of $G$ by trivially-acting $K$, and how a decomposition
arises in such orbifolds.  In particular, decomposition implements a 
restriction on nonperturbative sectors.  In an orbifold the nonperturbative
sectors are the twisted sectors, and in these orbifolds those twisted
sectors are restricted to those describing $G$ bundles satisfying a condition. 
The restriction is implemented physically by a sum over $G$ orbifolds,
namely the decomposition, realizing a `multiverse interference effect'
between the constituent $G$ orbifolds (`universes').
An important role in that decomposition is
played by discrete torsion, so in section~\ref{sect:3ddt} we review
three-dimensional analogues of discrete torsion, counted by
$H^3(G,U(1))$.  

In section~\ref{sect:2gp-orb} we turn to the main content
of this paper:  we define and study orbifolds by 2-group extensions
of ordinary (finite) groups $G$
by trivially-acting one-form symmetry groups $BK$.
Just as in two-dimensional cases,
the nonperturbative sectors correspond to $G$ bundles satisfying
a condition.  We argue that, also just as in two-dimensional cases,
that restriction implies (and is implemented by)
a decomposition of the three-dimensional theory,
with universes indexed by irreducible representations of $K$, which we
study explicitly in several examples.

In section~\ref{sect:sigma-2gerbe}
we interpret this structure formally
in terms of a sigma model whose target is
a 2-gerbe.  In section~\ref{sect:higher} we outline higher-dimensional
analogues and their interpretations.

In section~\ref{sect:cs} we briefly outline analogous decompositions
in Chern-Simons theories with gauged one-form symmetry group actions,
which will be further addressed in other work to appear.

In appendix~\ref{app:homotopy}, we give mathematically rigorous derivations
of statements about bundles of 2-groups.
In appendix~\ref{app:duality} we formally discuss decomposition as a duality
transform, as a type of Fourier transform.
Finally, in appendix~\ref{app:gpcohom} we collect some results on group
cohomology that are used in computations in the main text.

Higher-dimensional orbifolds have also been discussed in
e.g.~\cite{Freed:1991bn,Carqueville:2017aoe}.  
We believe our observations in this paper are novel.

\section{Review: decomposition in ordinary orbifolds}
\label{sect:ordinary-case}

In this section we will review decomposition of two-dimensional orbifolds
in which a central subgroup of the orbifold group acts trivially.
The fact that such orbifolds are equivalent to (`decompose into')
disjoint unions of
other theories was worked out in \cite{Hellerman:2006zs}; however,
our presentation
of the phenomenon here has not been previously published, and is the prototype
for our discussion of decomposition in 2-group orbifolds later.

Let $X$ be a space, and $G$ a finite group acting on $X$.
Let $\Gamma$ be a central extension of $G$ by a finite abelian
group $K$:
\begin{equation}
1 \: \longrightarrow \: K \: \longrightarrow \: \Gamma \: \longrightarrow \:
G \: \longrightarrow \: 1.
\end{equation}
Such extensions are classified by elements of $H^2(G,K)$.  Briefly,
the statement of decomposition here is that \cite{Hellerman:2006zs}
\begin{equation}  \label{eq:decomp:basic}
{\rm QFT}\left( [X/\Gamma] \right) \: = \: \coprod_{\rho \in \hat{K}} 
{\rm QFT}\left( [X/G]_{\rho(\omega)} \right),
\end{equation}
where $\hat{K}$ denotes irreducible representations of $K$,
and $\rho(\omega) \in H^2(G,U(1))$ 
is the image $\rho \circ \omega$ of the extension class $\omega$
under $\rho \in \hat{K}$.
(Decomposition is also defined for more general orbifolds
\cite{Hellerman:2006zs,Robbins:2020msp,Robbins:2021ibx},
but for our purposes in this paper, the special case of central
extensions above will suffice.)

Next, we establish this decomposition, by computing
partition functions.  First, recall that the extension
$\Gamma$ can be described set-wise as a product $G \times K$,
with product deformed by
an element $[\omega] \in H^2(G,K)$.  Let
$\gamma \in \Gamma$, and write $\Gamma$ set-wise as the product $G \times K$,
then the product in $\Gamma$ is defined by
\begin{equation}
\gamma_1 \gamma_2 \: = \: (g_1,k_1)\,(g_2,k_2) \: = \: \left( g_1 g_2, k_1 k_2 \omega(g_1, g_2) \right).
\end{equation}

In the partition function of a two-dimensional orbifold
$[X/\Gamma]$ on $T^2$, we sum over commuting pairs of group elements
in $\Gamma$, but clearly the condition for $\gamma_1$ and $\gamma_2$ to
commute is equivalent to $g_1$ commuting with $g_2$ and
\begin{equation}
\frac{ \omega(g_1, g_2) }{ \omega(g_2, g_1) } \: = \: 1.
\end{equation}

Define
\begin{equation}
\epsilon(g_1,g_2) \: = \: \frac{ \omega(g_1, g_2) }{ \omega(g_2, g_1) },
\end{equation}
then it is straightforward to demonstrate that
\begin{equation}
\epsilon(a,bc) \: = \: \epsilon(a,b) \epsilon(a,c),
\end{equation}
(and symmetrically,)
so as a consequence, $\epsilon$ is invariant under conjugation\footnote{
We restrict to the same $h$ on each input because $\epsilon$ is only
defined on commuting pairs.
}:
\begin{eqnarray}
\epsilon(h a h^{-1}, h b h^{-1}) 
& = &
\epsilon(h a h^{-1}, h) \, \epsilon(h a h^{-1}, b) \,
\epsilon(h a h^{-1}, h^{-1}),
\nonumber\\
& = &
\epsilon(h a h^{-1}, b),
\nonumber\\
& = &
\epsilon(h, b) \, \epsilon(a,b) \, \epsilon(h^{-1}, b),
\nonumber\\
& = &
\epsilon(a,b).
\end{eqnarray}
In particular, this descends to isomorphism classes of $G$ bundles,
which on $T^2$ are classified by Hom$(\pi_1(T^2),G)/G$.  We can view $\epsilon$ as assigning a phase to each such bundle.

Thus, the partition function of a two-dimensional $[X/\Gamma]$
orbifold looks like the partition function of a $[X/G]$ orbifold
but with a restriction on the allowed sectors.  We can implement that
restriction on allowed sectors by inserting an operator
\begin{equation}
\delta(\epsilon - 1) \: = \:
\frac{1}{|K|} \sum_{\rho \in \hat{K}} \epsilon_{\rho}(g_1, g_2),
\end{equation}
where $\epsilon_{\rho}$ is the image of $\omega(g_1, g_2)/\omega(g_2, g_1)$
under $\rho: K \rightarrow U(1)$.
This is the origin of decomposition \cite{Hellerman:2006zs}.

Now, let us assemble these pieces.  The partition function of a
$\Gamma$ orbifold on $T^2$ is, universally,
\begin{equation}
Z_{T^2}\left( [X/\Gamma] \right) \: = \: \frac{1}{|\Gamma|}
\sum_{\gamma \lambda = \lambda \gamma} Z(\gamma,\lambda),
\end{equation}
where the sum is over commuting pairs $\gamma, \lambda \in \Gamma$,
and $Z(\gamma,\lambda)$ is the contribution from a square with
sides identified by $\gamma$, $\lambda$ -- known as the twisted sectors or
partial traces.  In the present circumstances, since $K \subset \Gamma$
acts trivially,
\begin{equation}
Z(\gamma,\lambda) \: = \: Z(g,h)
\end{equation}
where $g = \pi(\gamma)$, $h = \pi(\lambda)$, for $\pi: \Gamma \rightarrow G$
the projection.  Taking this into account,
we then have
\begin{eqnarray}
Z_{T^2}\left( [X/\Gamma] \right) & = &
\frac{|K|^2}{|\Gamma|} \sum_{gh=hg, \epsilon=1} Z(g,h),
\nonumber\\
& = &
\frac{|K|^2}{|\Gamma|} \frac{|G|}{|K|} \sum_{\rho \in \hat{K}}
Z_{T^2}\left( [X/G]_{\rho(\omega)} \right),
\nonumber\\
& = &
\sum_{\rho \in \hat{K}}
Z_{T^2}\left( [X/G]_{\rho(\omega)} \right),
\end{eqnarray}
where
\begin{equation}
Z_{T^2}\left( [X/G]_{\rho(\omega)} \right) \: = \:
\frac{1}{|G|} \sum_{gh = hg} \epsilon_{\rho}(g,h) Z(g,h)
\end{equation}
is the partition function of the $G$ orbifold on $T^2$ with
discrete torsion $\rho(\omega) \in H^2(G,U(1))$.
Thus, we see that partition functions are consistent with the
prediction of decomposition~(\ref{eq:decomp:basic}).

In passing, note that in the case $G = {\mathbb Z}_2 = K$,
$H^2(G,K) = {\mathbb Z}_2$ (and hence has nontrivial elements),
but for all $[\omega] \in H^2(G,K)$, and all commuting pairs,
\begin{equation}
\frac{ \omega(g_1,g_2) }{ \omega(g_2,g_1) } \: = \: 1.
\end{equation}
Thus, triviality of the ratio of cocycles can happen even if $\omega$
is a nontrivial cohomology class.

Our analysis above was specific to the case that the worldsheet
is $T^2$, but it generalizes easily to other genus.
Before considering general genus, let us next walk through the case of
genus $2$.  Let $\gamma_i = (a_i, k_i) \in \Gamma$,
$\lambda_i = (b_i, z_i) \in \Gamma$, $i \in \{1, 2 \}$, obeying the
condition
\begin{equation} \label{eq:genus2-condition}
[ \gamma_1, \lambda_1] \, [\gamma_2, \lambda_2] \: = \: 1,
\end{equation}
for
\begin{equation}
[g,h] \: = \: g h g^{-1} h^{-1},
\end{equation}
and define
\begin{equation}
\xi_1 \: = \: [a_1,b_1] \: = \: a_1 b_1 a_1^{-1} b_1^{-1}.
\end{equation}
Then, using the fact that
\begin{equation}
\gamma_i^{-1} \: = \: \left(a_i^{-1}, k_i^{-1} \omega(a_i,a_i^{-1})^{-1}
 \right),
\: \: \:
\lambda_i^{-1} \: = \: \left( b_i^{-1}, z_i^{-1} \omega(b_i, b_i^{-1})^{-1}
\right),
\end{equation}
it is straightforward to compute that   
\begin{eqnarray}
[\gamma_1, \lambda_1] & = &
\left( [a_1, b_1], 
\omega(a_1,b_1) \, \omega(a_1 b_1, a_1^{-1}) \,
 \omega(a_1 b_1 a_1^{-1}, b_1^{-1}) \,
\right. \nonumber \\
& & \hspace*{1.1in} \left. \cdot
\omega(a_1,a_1^{-1})^{-1} \, \omega(b_1, b_1^{-1})^{-1}
\right),
\end{eqnarray}
\begin{eqnarray}
[\gamma_1,\lambda_1] \, [\gamma_2, \lambda_2] 
& = &
\left( [a_1, b_1] [a_2, b_2], 
\omega(a_1,b_1) \, \omega(a_1 b_1, a_1^{-1}) \,
 \omega(a_1 b_1 a_1^{-1}, b_1^{-1}) \,
\right. \nonumber \\
& & \hspace*{1.1in} \left. \cdot
\omega(\xi_1,a_2) \, \omega(\xi_1 a_2, b_2) \, 
\omega(\xi_1 a_2 b_2, a_2^{-1}) \,
\right. \nonumber \\
& & \hspace*{1.1in} \left. \cdot
\omega(\xi_1 a_2 b_2 a_2^{-1}, b_2^{-1})
\right. \nonumber \\
& & \hspace*{1.1in} \left. \cdot
\omega(a_1, a_1^{-1})^{-1} \, \omega(a_2, a_2^{-1})^{-1} \,
\right. \nonumber \\
& & \hspace*{1.1in} \left. \cdot
\omega(b_1, b_1^{-1})^{-1} \, \omega(b_2, b_2^{-1})^{-1}
\right),
\end{eqnarray}
so we see that the closure condition~(\ref{eq:genus2-condition}) holds
if and only if
both
\begin{equation}
[a_1, b_1] \, [a_2, b_2] \: = \: 1
\end{equation}
and
\begin{eqnarray}
1 & = &
\omega(a_1,b_1) \, \omega(a_1 b_1, a_1^{-1}) \,
 \omega(a_1 b_1 a_1^{-1}, b_1^{-1}) \,
\nonumber \\
& & \hspace*{0.5in} \cdot
\omega(\xi_1,a_2) \, \omega(\xi_1 a_2, b_2) \, 
\omega(\xi_1 a_2 b_2, a_2^{-1}) \,
\omega(\xi_1 a_2 b_2 a_2^{-1}, b_2^{-1})
\nonumber \\
& & \hspace*{0.5in} \cdot
\omega(a_1, a_1^{-1})^{-1} \, \omega(a_2, a_2^{-1})^{-1} \,
\omega(b_1, b_1^{-1})^{-1} \, \omega(b_2, b_2^{-1})^{-1}.
\end{eqnarray}

Next, we generalize to arbitrary genus.
Consider a Riemann surface of genus $g$,
with boundary conditions determined by $\gamma_i = (a_i, k_i) \in \Gamma$,
$\lambda_i = (b_i, z_i) \in \Gamma$, $i \in \{1, \cdots, g\}$.
Define $\xi_i = [a_i, b_i]$,
and
\begin{equation}
        X \: = \: \left[ \prod_i \omega(a_i, a_i^{-1})
        \prod_i \omega(b_i,b_i^{-1}) \right]^{-1}.
\end{equation}
The condition that the group elements must obey to define boundary
conditions on the Riemann surface is that
\begin{equation}
[ \gamma_1, \lambda_1] \, [\gamma_2, \lambda_2] \cdots
[\gamma_g, \lambda_g] \: = \: 1,
\end{equation}
which implies that
\begin{equation}
[ a_1, b_1] \, [a_2, b_2] \cdots [a_g, b_g] \: = \: 1
\end{equation}
(which are required for $a_i, b_i \in G$ to close on the Riemann surface)
as well as
\begin{equation}
\epsilon(a_i, b_i) \: = \: 1
\end{equation}
for
\begin{eqnarray}
\epsilon(a_i, b_i) 
& \equiv &
X
        \omega(a_1, b_1) \omega(a_1 b_1, a_1^{-1})
        \omega(a_1 b_1 a_1^{-1}, b_1^{-1})
        \omega(\xi_1, a_2) \omega(\xi_1 a_2, b_2)
        \omega(\xi_1 a_2 b_2, a_2^{-1})
\nonumber \\
& &  \hspace*{1in} \cdot
        \omega(\xi_1 a_2 b_2 a_2^{-1}, b_2^{-1})
        \omega(\xi_1 \xi_2, a_3) \cdots
        \omega(\xi_1 \cdots \xi_{g-1} a_g b_g a_g^{-1}, b_g^{-1}).
\nonumber
\end{eqnarray}
(This can be obtained either by direct multiplication or by
triangulating the Riemann surface into simplices and associating
a factor of $\omega$ with each simplex, as in \cite{Aspinwall:2000xv}.)
Thus, as before, the data required to define a $\Gamma$
orbifold on a genus $g$ Riemann surface
is a restriction on the combinatorial data used to define
a $G$ orbifold on the same Riemann surface, a restriction of the
form $\epsilon(a_i, b_i) = 1$.  As for $T^2$, we can implement that
restriction by inserting a projection operator $\Pi$, of the same form
as before, with $\epsilon_{\rho}$ that are the image of the
genus-$g$ $\epsilon$ under an irreducible representation $\rho$.
The resulting phases are the same as the phases defining discrete torsion
on a genus $g$ Riemann surface (see \cite[equ'n (15)]{Aspinwall:2000xv},
\cite{Bantay:2000eq}),
again for discrete torsion given by the image of $H^2(G,K)$ under
the irreducible representation $\rho: K \rightarrow U(1)$.
Thus, we see the story for $T^2$ generalizes immediately to other
Riemann surfaces.

For later use, we note that the discrete torsion here can equivalently
be understood as a coupling to a discrete theta angle, defined by
a characteristic class $x^* \omega$, for $x: \Sigma \rightarrow BG$
a map defining the twisted sector, in the notation of
appendix~\ref{app:homotopy}.  One can rewrite such a discrete theta angle
coupling
\begin{equation}
\int_{\Sigma}
\langle \rho, x^* \omega \rangle
\end{equation}
as a discrete torsion phase by triangulating the Riemann surface
$\Sigma$ and associating phases to each simplex as reviewed above and in
\cite{Aspinwall:2000xv,Dijkgraaf:1989pz}.

So far we have considered central extensions.  Decomposition also exists
for orbifolds by non-central extensions, 
see e.g.~\cite{Hellerman:2006zs,Robbins:2020msp};
however, its form is more complex.  In this paper we focus on
(analogues of) central extensions.

\section{Three-dimensional analogues of discrete torsion}
\label{sect:3ddt}

We have seen that
two-dimensional orbifolds with trivially-acting subgroups decompose
into disjoint unions of orbifolds with discrete torsion, a modular-invariant
phase factor \cite{Hellerman:2006zs,Robbins:2020msp}.  
Similarly, the three-dimensional version of
decomposition will also generate theories twisted by a three-dimensional
version of discrete torsion.
Such analogues of discrete torsion were studied in
\cite{Dijkgraaf:1989pz} in the special case of orbifolds of points
(forming Dijkgraaf-Witten theory), and more generally in
\cite{Sharpe:2000qt}.  In this section,
we briefly review those constructions here, in both ordinary orbifolds
and in orientifolds, to set up their appearance
in three-dimensional versions of decomposition.

\subsection{Ordinary orbifolds}

First, recall that
in two dimensions, discrete torsion in a $G$ orbifold is classified
by group cohomology, specifically $H^2(G,U(1))$ with a trivial action
on the coefficients.  Similarly, in three
dimensions \cite{Dijkgraaf:1989pz,Sharpe:2000qt}, 
the analogue of discrete torsion
in a $G$ orbifold
is classified by $H^3(G,U(1))$, again with a trivial action on the coefficients.

Furthermore,
given $[\omega] \in H^2(G,U(1))$, one can derive
coboundary-invariant phases that weight Riemann surfaces.  
For example, on $T^2$, a twisted sector is
defined by two commuting elements $g, h \in G$, 
and the corresponding
coboundary-invariant phase is
\begin{equation}
\frac{ \omega(g,h) }{ \omega(h,g) }.
\end{equation}
Analogous expressions on higher-genus Riemann surfaces can be
found in \cite{Aspinwall:2000xv}.

Analogous constructions exist in three dimensions, which use
$[\omega] \in H^3(G,U(1))$ to assign a coboundary-invariant
phase to three-manifolds.  
One construction \cite{Dijkgraaf:1989pz} proceeds as follows.
given a three-manifold $Y$, we pick a triangulation by simplices,
and associate to each simplex a cocycle.  We then take an alternating
product of those associated cocycles (with exponent determined by
orientation) to form a coboundary-invariant phase.
For example, the triangulation of a cube into six simplices can be
visualized by viewing the cube along a line through two corners, as
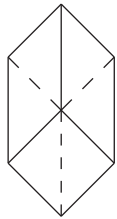
\begin{center}
\begin{picture}(40,80)(0,0)
\Line(20,40)(20,80)  \Line(20,40)(0,20)  \Line(20,40)(40,20)
\DashLine(20,40)(20,0){5}  \DashLine(20,40)(0,60){5}
\DashLine(20,40)(40,60){5}
\Line(0,20)(0,60) \Line(40,20)(40,60)
\Line(0,60)(20,80)  \Line(40,60)(20,80)
\Line(20,0)(0,20)  \Line(20,0)(40,20)
\end{picture}
\end{center}
and then taking the tetrahedra cut out by the six interior lines 
projected through the cube in the figure
above.
See also \cite{Sharpe:2000qt}
for an alternative construction in terms of $C$ field holonomlies.

For example, on $T^3$, a twisted sector is defined by three commuting group
elements $g_1, g_2, g_3$, 
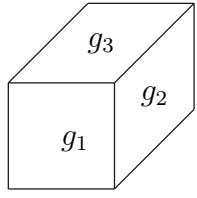
\begin{center}
\begin{picture}(70,70)(0,0)
\Line(0,0)(40,0)  \Line(0,0)(0,40)
\Line(40,0)(40,40) \Line(0,40)(40,40)
\Line(0,40)(30,70) \Line(30,70)(70,70)
\Line(40,40)(70,70)  \Line(40,0)(70,30)
\Line(70,30)(70,70)
\Text(20,18)[l]{$g_1$}
\Text(30,55)[l]{$g_3$}
\Text(50,35)[l]{$g_2$}
\end{picture}
\end{center}
and here one
multiplies $Z(g_1, g_2, g_3)$ by the phase
\cite[equ'n (6.35)]{Dijkgraaf:1989pz}, \cite{Sharpe:2000qt}
\begin{equation}  \label{eq:dw-phases}
\epsilon_3(g_1,g_2,g_3) \: = \:
\frac{ \omega(g_1, g_2, g_3) }{ \omega(g_2, g_1, g_3) }
\frac{ \omega(g_3, g_1, g_2) }{ \omega(g_3, g_2, g_1) }
\frac{ \omega(g_2, g_3, g_1) }{ \omega(g_1, g_3, g_2) }.
\end{equation}
corresponding to $[\omega] \in H^3(G,U(1))$.
As noted in \cite[footnote 5]{Dijkgraaf:1989pz}, perhaps the
simplest example in which this phase is nontrivial is the
group $G = ( {\mathbb Z}_2 )^3$.

As discussed in \cite{Dijkgraaf:1989pz,Sharpe:2000qt},
this phase factor is invariant
under both coboundaries as well as
$SL(3,{\mathbb Z})$ transformations of $T^3$, just as the
discrete torsion phase factor is invariant under both coboundaries as
well as $SL(2,{\mathbb Z})$
transformations of $T^2$.

For another example, consider $S^1 \times \Sigma$ for $\Sigma$ a genus-two
surface.  Here,
the associated phase is
\begin{eqnarray}
\xi_2 & = & 
\frac{\omega(a_1, b_1, g) }{ \omega(\gamma b_1, a_1, g) \, \omega(\gamma, b_1, g) }
\frac{ \omega(\gamma, a_2, g) \, \omega(\gamma a_2, b_2, g) }{ \omega(b_2, a_2, g)}
\nonumber \\
&  & \hspace*{0.25in}  \cdot
\frac{ \omega(\gamma b_1, g, a_1) \, \omega(\gamma, g, b_1) }{ \omega(a_1, g, b_1)}
\frac{\omega(b_2, g, a_2)}{\omega(\gamma, g, a_2) \, \omega(\gamma a_2, g, b_2)}
\nonumber \\
&  & \hspace*{0.25in} \cdot
\frac{\omega(g, a_1, b_1)}{\omega(g, \gamma b_1, a_1) \, \omega(g,\gamma, b_1)}
\frac{\omega(g,\gamma,a_2) \, \omega(g,\gamma a_2, b_2)}{\omega(g,b_2,a_2)}
\end{eqnarray}
where
\begin{equation}
\gamma \: = \: a_1 b_1 a_1^{-1} b_1^{-1},
\: \: \:
\gamma a_2 b_2 a_2^{-1} b_2^{-1} \: = \: 1.
\end{equation}
and $g$ commutes with all $a_i$, $b_i$.
It can be shown that this expression is invariant under coboundaries.

This expression is motivated by the two-dimensional genus-two phase
\cite[equ'n (15)]{Aspinwall:2000xv}
\begin{equation}
\frac{\omega(a_1,b_1)}{\omega(\gamma b_1, a_1) \, \omega(\gamma, b_1)}
\frac{\omega(\gamma, a_2) \, \omega(\gamma a_2, b_2)}{\omega(b_2, a_2)}.
\end{equation}
Also, in the special case that $\gamma = 1$, it correctly factorizes into
the product of two $T^3$ phases:
\begin{equation}
\xi_2 \: = \: \frac{\omega(a_1, b_1, g)}{\omega(b_1, a_1, g)}
\frac{\omega(b_1, g,a_1)}{\omega(a_1, g, b_1)}
\frac{\omega(g, a_1, b_1)}{\omega(g, b_1, a_1)}
\cdot
\frac{\omega(a_2,b_2,g)}{\omega(b_2,a_2,g)}
\frac{\omega(b_2,g,a_2)}{\omega(a_2,g,b_2)}
\frac{\omega(g,a_2,b_2)}{\omega(g,b_2,a_2)},
\end{equation}
where without loss of generality we assume tha the cocycle $\omega$ is
normalized (so that $\omega = 1$ if any of its arguments is the identity).

In fact, it is also straightforward to conjecture the corresponding
phase factor for $S^1 \times \Sigma_h$ for $\Sigma_h$ a genus-$h$
Riemann surface.  Following \cite{Aspinwall:2000xv}, define
\begin{equation}
\gamma_i \: = \: a_i b_i a_i^{-1} b_i^{-1},
\: \: \:
\zeta_i \: = \: \gamma_1 \gamma_2 \cdots \gamma_{i-1},
\end{equation}
then the two-dimensional discrete torsion phase is
\cite[equ'n (15)]{Aspinwall:2000xv}
\begin{equation}
\xi_h \: = \:
\frac{ \omega(a_1,b_1) }{ \omega(\gamma_1 b_1, a_1) \, \omega(\gamma_1,b_1) }
\left( \prod_{i=2}^{h-1} 
\frac{ \omega(\zeta_i, a_i) \, \omega(\zeta_i a_i, b_i) }{
\omega(\zeta_i \gamma_i b_i, a_i) \, \omega(\zeta_i \gamma_i, b_i) }
\right)
\frac{
\omega(\zeta_h, a_h) \, \omega(\zeta_h a_h, b_h) }{ \omega(b_h, a_h) }
\end{equation}
and we conjecture that the analogous three-dimensional phase on
$S^1 \times \Sigma_h$ is
\begin{eqnarray}
\xi_h & = &
\frac{ \omega(a_1,b_1,g) }{ \omega(\gamma_1 b_1, a_1,g) \, \omega(\gamma_1,b_1,g) }
\frac{ \omega(\gamma_1 b_1, g, a_1) \, \omega(\gamma_1, g, b_1) }{
\omega(a_1,g,b_1) }
\frac{ \omega(g, a_1,b_1) }{ \omega(g, \gamma_1 b_1, a_1) \, \omega(g, \gamma_1,b_1) }
\nonumber \\
& & \hspace*{0.25in} \cdot
\Biggl( 
\prod_{i=2}^{h-1} 
\frac{ \omega(\zeta_i, a_i,g) \, \omega(\zeta_i a_i, b_i,g) }{
\omega(\zeta_i \gamma_i b_i, a_i,g) \, \omega(\zeta_i \gamma_i, b_i,g) }
\frac{
\omega(\zeta_i \gamma_i b_i, g, a_i) \, \omega(\zeta_i \gamma_i, g, b_i) }
{ \omega(\zeta_i, g, a_i) \, \omega(\zeta_i a_i, g, b_i) }
\nonumber \\
& & \hspace*{2.0in} \cdot
\frac{ \omega(g, \zeta_i, a_i) \, \omega(g, \zeta_i a_i, b_i) }{
\omega(g, \zeta_i \gamma_i b_i, a_i) \, \omega(g, \zeta_i \gamma_i, b_i) }
\Biggr)
\nonumber \\
& & \hspace*{0.25in} \cdot
\frac{
\omega(\zeta_h, a_h,g) \, \omega(\zeta_h a_h, b_h,g) }{ \omega(b_h, a_h,g) }
\frac{ \omega(b_h, g, a_h) }{
\omega(\zeta_h, g, a_h) \, \omega(\zeta_h a_h, g, b_h) }
\nonumber \\
& & \hspace*{2.5in} \cdot
\frac{
\omega(g, \zeta_h, a_h) \, \omega(g, \zeta_h a_h, b_h) }{ \omega(g, b_h, a_h) }
\end{eqnarray}
where $g \in G$ commutes with all $a_i$, $b_i$.

In two dimensions, discrete torsion phases obey multiloop factorization
(target space unitarity), which is the following
constraint.  If $\Sigma$ is any Riemann surface, corresponding to a twisted
sector of some orbifold, and $\Sigma$ can degenerate into
a product of $\Sigma_1$ and $\Sigma_2$ connected at one point
(compatibly with the orbifold structure,
in the sense that there are no twist fields at the connection),
then the phase associated to $\Sigma$ must equal the
product of the phases associated to $\Sigma_1$ and $\Sigma_2$.

In two dimensions, for the genus-one phase
\begin{equation}
\epsilon_2(g,h) \: = \: \frac{\omega(g,h)}{\omega(h,g)},
\end{equation}
this is the property \cite[equ'n (42)]{Vafa:1986wx}
\begin{equation} \label{eq:eps2-hom}
\epsilon_2(x,ab) \: = \: \epsilon_2(x,a) \, \epsilon_2(x,b),
\end{equation}
which can be demonstrated simply using
\begin{equation}
\frac{ (d \omega)(x,a,b) \, (d \omega)(a,b,x) }{ (d \omega)(a,x,b) }
\: = \:
\frac{ \epsilon_2(x,ab) }{ \epsilon_2(x,a) \, \epsilon_2(x,b) }
\end{equation}
for $x$, $a$, $b$ all mutually commuting.
When combined with the fact that $\epsilon_2(1,-) = \epsilon_2(-,1) = 1$,
we see this means
that $\epsilon_2$ is a bihomomorphism from
commuting pairs in $G$ to $U(1)$.

In three dimensions, there is a simple analogue of multiloop factorization:
if a three-manifold $S^1 \times \Sigma$ can degenerate into
$S^1 \times ( \Sigma_1 \coprod \Sigma_2)$, the the phase assigned
to $S^1 \times \Sigma$ must match the product of the phases assigned
to $S^1 \times \Sigma_1$, $S^1 \times \Sigma_2$.
On such grounds, one then expects
\begin{equation}  \label{eq:eps3-hom}
\epsilon_3(x,y,ab) \: = \: \epsilon_3(x,y,a) \epsilon_3(x,y,b).
\end{equation}
In fact, it is straightforward to check that this is
a consequence of the identity
\begin{equation}
\frac{ (d\omega)(y,x,a,b) }{ (d\omega)(x,y,a,b) }
\frac{ (d\omega)(a,b,y,x) }{ (d\omega)(a,b,x,y) }
\frac{ (d\omega)(y,a,b,x) }{ (d\omega)(x,a,b,y) }
\frac{ (d\omega)(x,a,y,b) }{ (d\omega)(y,a,x,b) }
\frac{ (d\omega)(a,x,b,y) }{ (d\omega)(a,y,b,x) }
\frac{ (d\omega)(a,y,x,b) }{ (d\omega)(a,x,y,b) }
\: = \: 1.
\end{equation}
(See also \cite[section 6]{Dijkgraaf:1989pz}, where a different argument
is given for the same result.)

One can use multiloop factorization to argue that discrete torsion(-like)
phases descend to conjugacy classes.  For example, in the case of
the genus-one phase $\epsilon_2$, from~(\ref{eq:eps2-hom}), it is easy
to show that\footnote{In fact, formally both this expression and its
three-dimensional analogue appear to generalize to independent conjugatiosn
on the parameters, as
\begin{equation}
\epsilon_2(a g a^{-1}, b h b^{-1}) \: = \:
\epsilon_2(g,h).
\end{equation}
However, $\epsilon_2(g,h)$ is only defined for commuting $g$, $h$,
so we restrict to the case $a=b$.  Identical remarks apply to
$\epsilon_3$.
}
\begin{eqnarray}
\epsilon_2(a g a^{-1}, a h a^{-1}) 
& = &
\epsilon_2(a g a^{-1}, a) \, 
\epsilon_2(a g a^{-1}, h)\,
\epsilon_2(a g a^{-1}, a^{-1})
\: = \: \epsilon_2(a g a^{-1}, h),
\nonumber\\
& = &
\epsilon_2(a,h) \, \epsilon_2(g, h) \, \epsilon_2(a^{-1}, h),
\nonumber\\
& = &
\epsilon_2(g,h).
\end{eqnarray}
Computing in exactly the same fashion, one can use~(\ref{eq:eps3-hom})
to show that
\begin{equation}
\epsilon_3(a g a^{-1}, a h a^{-1}, a k a^{-1}) \: = \:
\epsilon_3(g,h,k).
\end{equation}

\subsection{Manifolds with boundaries}

For completeness, let us also quickly outline the case of manifolds with
boundary, that will be of use in our subsequent works.

Let's begin with an overview of how this works in two-dimensional
theories and ordinary discrete torsion, before describing an example
in three dimensions.

Consider a genus-one correlation function in an orbifold $[X/G]$
with a single insertion
of an operator associated to $g \in G$.  In effect, we have a $T^2$ with
a puncture corresponding to $g$.  If we let $a, b \in G$
denote group elements corresponding to the usual $T^2$ boundary conditions,
then
we can sketch the construction of the punctured torus as in the
diagram below:
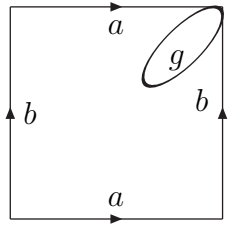
\begin{center}
\begin{picture}(80,80)(0,0)
\ArrowLine(0,0)(80,0)
\ArrowLine(0,0)(0,80)
\ArrowLine(80,0)(80,80)
\ArrowLine(0,80)(80,80)
\Oval(65,65)(7,20)(45)
\Text(40,5)[b]{$a$}
\Text(5,40)[l]{$b$}
\Text(40,75)[t]{$a$}
\Text(75,45)[r]{$b$}
\Text(60,60)[l]{$g$}
\end{picture}
\end{center}
with a hole cut out in the upper right corner, or equivalently,
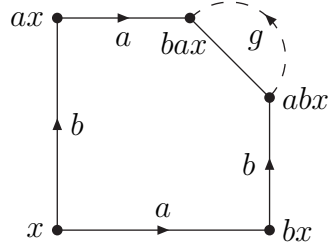
\begin{center}
\begin{picture}(80,80)(0,0)
\ArrowLine(0,0)(80,0)
\ArrowLine(0,0)(0,80)
\ArrowLine(80,0)(80,50)
\ArrowLine(0,80)(50,80)
\Line(50,80)(80,50)
\Text(40,5)[b]{$a$}
\Text(5,40)[l]{$b$}
\Text(25,75)[t]{$a$}
\Text(75,25)[r]{$b$}
\Vertex(0,0){2} \Text(-5,0)[r]{$x$}
\Vertex(80,0){2} \Text(85,0)[l]{$bx$}
\Vertex(0,80){2} \Text(-5,80)[r]{$ax$}
\Vertex(50,80){2}  \Text(48,75)[t]{$bax$} 
\Vertex(80,50){2}  \Text(85,50)[l]{$abx$} 
\Text(72,72)[l]{$g$}
\DashArrowArc(65,65)(21,-45,135){5}
\end{picture}
\end{center}
In the presence of the puncture, $a$ and $b$ no longer commute,
but instead obey
\begin{equation} \label{eq:t2-punc}
abg \: = \: ba.
\end{equation}

Alternatively, we can say that if $\Sigma$ is a punctured $T^2$, 
then to specify an element of Hom$(\pi_1(\Sigma),G)$ 
we can first assign a group element $g$ to the loop circling the puncture, 
and then the group elements $a$ and $b$ assigned to the non-contractible cycles 
of the torus will need to satisfy~(\ref{eq:t2-punc}), 
since the cycle associated to $a^{-1}b^{-1}ab$ is homotopic to the cycle circling the puncture.  
Then bundles on $\Sigma$ are classified by Hom$(\pi_1(\Sigma),G)/G$, as usual.

One more perspective comes from consideration of topological defect lines.  
The $a$ and $b$ twists on the $T^2$ are implemented by wrapping $a$ and $b$ 
lines around the cycles.  
Saying that our inserted operator is associated to $g$ is equivalent to saying 
that it sits at the end of a defect line labeled by $g$.  
The other end of that line must terminate somewhere 
on the first two defect lines.  
The simplest possibility is to connect everything 
at a single junction of degree five.  In order for that junction 
to remain topological (i.e.~to avoid an extra non-topological insertion), 
we need the cyclic product of lines coming in to give the identity, 
which again leads to~(\ref{eq:t2-punc}).

In any event, 
the discrete torsion phase assigned to a punctured $T^2$ is not the same
as that assigned
to $T^2$ itself -- the contribution to the boundary conditions from
the puncture modifies the phase.
Applying methods of \cite{Aspinwall:2000xv}, we see that the phase associated
to this diagram is
\begin{equation}
\xi_{1,1} \: \equiv \: \frac{ \omega(a,b) }{ \omega(b,a) } \, \omega(ab,g).
\end{equation}
Now, if we add a coboundary $\alpha$, this phase changes:
\begin{equation}
\xi_{1,1} \: \mapsto \: \xi_{1,1} \, \alpha(g).
\end{equation}
This is not quite invariant under coboundaries; however,
the coboundary $\alpha(g)$ can be absorbed into the operator at the
puncture, so taking that into account, the phase is well-defined.
Proceeding in this fashion, one is led to correlation functions,
see e.g.~\cite{Ramgoolam:2022xfk} 
for examples in the case of orbifolds of a point
(Dijkgraaf-Witten theory).

Now, let us turn to three-dimensional analogues.
We consider a $T^3$ with a hole, of boundary $T^2$.
If we let the three primary sides be related by $g_1$, $g_2$, $g_3$,
and the new edge defining the hole by $k$, then graphically,
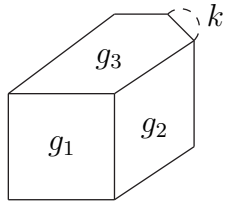
\begin{center}
\begin{picture}(70,70)(0,0)
\Line(0,0)(40,0)  \Line(0,0)(0,40)
\Line(40,0)(40,40) \Line(0,40)(40,40)
\Line(0,40)(40,70) \Line(40,70)(60,70)  \Line(60,70)(70,60)
\Line(40,40)(70,60)  \Line(40,0)(70,20) 
\Line(70,20)(70,60)
\Text(15,20)[l]{$g_1$}
\Text(33,53)[l]{$g_3$}
\Text(50,27)[l]{$g_2$}
\DashCArc(65,65)(7,-45,135){3}  \Text(75,70)[l]{$k$}
\end{picture}
\end{center}
The cut-out corner, seen edge-on, is the square 
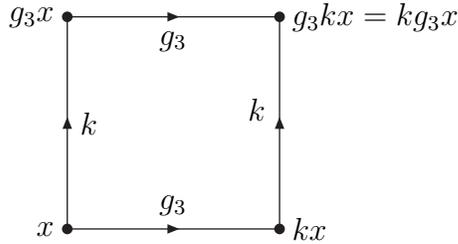
\begin{center}
\begin{picture}(80,80)(0,0)
\ArrowLine(0,0)(80,0)
\ArrowLine(0,0)(0,80)
\ArrowLine(80,0)(80,80)
\ArrowLine(0,80)(80,80)
\Text(40,5)[b]{$g_3$}
\Text(5,40)[l]{$k$}
\Text(40,75)[t]{$g_3$}
\Text(75,45)[r]{$k$}
\Vertex(0,0){2} \Text(-5,0)[r]{$x$}
\Vertex(0,80){2} \Text(-5,80)[r]{$g_3 x$}
\Vertex(80,0){2} \Text(85,0)[l]{$kx$}
\Vertex(80,80){2}  \Text(85,80)[l]{$g_3 k x = k g_3 x$}
\end{picture}
\end{center}
In order for the diagram to close, the four group elements
$g_1, g_2, g_3, k \in G$ obey
\begin{equation}
g_1 g_3 \: = \: g_3 g_1, \: \: \:
g_2 g_3 \: = \: g_3 g_2, \: \: \:
g_1 g_2 k \: = \: g_2 g_1, \: \: \:
g_3 k \: = \: k g_3.
\end{equation}
Applying the same methods as \cite{Sharpe:2000qt}, we find that the
phase factor associated with this diagram is
\begin{equation}
\frac{ \omega(g_1, g_2, g_3) }{ \omega(g_2, g_1, g_3) }
\frac{ \omega(g_3, g_1, g_2) }{ \omega(g_1, g_3, g_2) }
\frac{ \omega(g_2, g_3, g_1) }{ \omega(g_3, g_2, g_1) }
\frac{ \omega(g_1 g_2, k, g_3) \, \omega(g_3, g_1 g_2, k) }{
\omega(g_1 g_2, g_3, k) }.
\end{equation}
As for the $T^2$ with boundary, this is not quite coboundary-invariant,
but rather picks up a phase
\begin{equation}
\frac{ \alpha(k,g_3) }{ \alpha(g_3, k) },
\end{equation}
which has the same appearance as the phase one would assign to a 
$T^2$ with the same boundary conditions.
We interpret this as before, as a contribution that would be absorbed by
a defect inserted at the puncture, precisely in the spirit of
anomaly inflow (see e.g.~\cite{Callan:1984sa}).
It is also extremely reminiscent of the relationship between
three-dimensional Chern-Simons theories and WZW models on boundaries,
see e.g.~\cite{Moore:1989yh}.

\subsection{Orientifolds}

Now, consider the case that a subgroup of the orbifold group $G$ acts,
in part, by reversing orientations, to form an orientifold.  
$C$ fields on orientifolds were analyzed in
\cite[section 6]{Sharpe:2009hr}, 
in the same pattern as in \cite{Sharpe:2000qt} for $C$ fields on ordinary
orbifolds and
\cite{Sharpe:2009hr} for $B$ fields on
orientifolds.  
Briefly, the conclusion was that the analogue of $C$ field discrete torsion
on orientifolds is counted by $H^3(G,U(1))$ with a nontrivial action on
the coefficients, encoded in a homomorphism $\epsilon: G \rightarrow 
{\mathbb Z}_2$ expressing whether a given element acts trivially.

One example discussed in \cite[section 6.2]{Sharpe:2009hr}
is a cube, with sides identified by three group elements
$g_1, g_2, g_3 \in G$, in which one of the group elements reverses the
orientation.  
The three group elements must be related by
\begin{equation}
g_2 g_3 \: = \: g_3 g_2, \: \: \:
g_1 g_3 \: = \: g_3 g_1, \: \: \:
g_1 \: = \: g_2 g_1 g_2,
\end{equation}
where of the three, $g_1$ reverses orientation, but the other two do not.
It was argued there that the corresponding
partition function phase factor is
\begin{equation}
\frac{ 
\omega(g_1, g_2^{-1}, g_3) \, \omega(g_2, g_3, g_1) \, \omega(g_3, g_1, g_2^{-1})
}{
\omega(g_2, g_1, g_3) \, \omega(g_1, g_3, g_2^{-1}) \, \omega(g_3, g_2, g_1)
}
\frac{
\omega(g_3, g_2, g_2^{-1}) \, \omega(g_2, g_2^{-1})
}{
\omega(g_2, g_3, g_2^{-1})
},
\end{equation}
which is invariant under coboundaries.

As another example, consider $S^1 \times {\mathbb R} {\mathbb P}^2$.
Let $g \in G$ be orientation-reversing, with $g^2 = 1$,
and $h \in G$ any other element that commutes with $g$.
It was argued in \cite[section 5.2]{Sharpe:2009hr}
that on the real projective plane ${\mathbb R} {\mathbb P}^2$,
with sides identified by $g$, the discrete torsion phase is
$\omega(g,g)$, which for $g^2 = 1$ is easily checked to be coboundary-invariant.
For $S^1 \times {\mathbb R} {\mathbb P}^2$, where the real projective
plane is again constructed with $g$, 
the $C$ field discrete torsion phase can be shown to be
\begin{equation}
\frac{
\omega(g,g,h)  \,  \omega(h,g,g)
}{
\omega(g,h,g)
},
\end{equation}
which is easily checked to be coboundary-invariant.

\section{Orbifolds by 2-groups}
\label{sect:2gp-orb}

In this section we will discuss three-dimensional\footnote{
As also noted in the introduction, throughout we have in mind
effective field theories as prototypes, though our methods also
apply more generally.
} 
orbifolds by 2-group extensions.
We saw in section~\ref{sect:ordinary-case} 
that an ordinary orbfiold by a central group
extension of $G$ by trivially-acting $K$ involves a restriction on
permitted $G$ bundles, which is implemented by the sum over universes.
We shall see an analogous structure here:  the 2-group orbifold will
involve a restriction on permitted $G$ bundles, which is implemented by
a sum over universes.
In this fashion we will derive a decomposition, which we will check
in examples.

In passing, we should mention that just as Dijkgraaf-Witten topological
field theory \cite{Dijkgraaf:1989pz} 
can be interpreted as an orbifold of a point,
at least naively the Yetter model \cite{yetter,br,mackaay,porter,fmp,fhlt}
appears to be interpretable 
as a 2-group (or higher) group orbifold of a point.
We will not pursue that in this paper, however.

\subsection{General aspects}
\label{sect:orb-2gp-genl}

\subsubsection{Notions of 2-groups and their gauging}

A 2-group is, roughly, a group in which associativity holds only up to
isomorphisms.
In this section we will outline orbifolds by 2-groups, and their
decomposition.

Briefly, from \cite[section 8.3]{baezlauda},
given a group $G$ and an abelian group $K$, to specify a (coherent) 2-group
one specifies an action $\alpha: G \rightarrow {\rm Aut}(K)$
plus an element of $H^3(G,K)$, where the group
cohomology is defined
with the action of $G$
on $K$ given by $\alpha$.  In this section we will restrict
to the analogue of a central extension, for which the map
$\alpha$ is trivial, and for which $H^3(G,K)$ is defined with trivial
action on the coefficients.

We will describe 2-groups as extensions of the form
\begin{equation}
1 \: \longrightarrow \: BK \: \longrightarrow \: \tilde{\Gamma}
\: \longrightarrow \: G \: \longrightarrow \: 1,
\end{equation}
for finite abelian $K$.  These are
classified by $[\omega] \in H^3(G,K)$.

In broad brushstrokes, to gauge a 2-group $\tilde{\Gamma}$ means that
the path integral
\begin{itemize}
\item sums over $K$ gerbes, and within that, for each $K$ gerbe,
\item sums over $G$ bundles twisted by the action of the $K$ gerbe,
in the sense of e.g.~\cite{Witten:1998cd}.
\end{itemize}
(In general there may also be other mutual twistings, 
as in e.g.~\cite[equ'ns (1.10), (1.14)]{Cordova:2020tij},
implementing a Green-Schwarz mechanism,
in which case one would not have for example precisely a path integral
over ordinary $K$ gerbes, but rather over slightly different objects
forming a torsor under $K$ gerbes.)
Examples in which the $K$ gerbe acts nontrivially (via the action of 
$BK$ on line operators, for example) include the gauging
that arose in \cite{Sharpe:2019ddn}, and also in discussions of
gauging $BK$ in Chern-Simons theories for $K$ the center of the
gauge group.

In this paper, we will be focused on the case in which the one-form
symmetry group being gauged acts completely trivially on the 
three-dimensional theory,
meaning that line operators are invariant\footnote{
For example, consider $SU(2)$ Chern-Simons theory in three dimensions.
This has a $B {\mathbb Z}_2$ one-form symmetry, inherited from the
center of $SU(2)$.  However, that one-form symmetry multiplies Wilson
lines by phases, and so we would not characterize $SU(2)$ Chern-Simons
as invariant under this $B {\mathbb Z}_2$.  One could in principle
consider a different $B {\mathbb Z}_2$, unrelated to the central
${\mathbb Z}_2$, which leaves all Wilson lines invariant.
In that case, that $B {\mathbb Z}_2$ could be said to act
trivially.
} under $BK$, meaning for example that associated line
operators have no braiding with one another or with any of the line operators
in the theory being gauged.  In this case,
relevant for us in this paper, we will see that
gauging a 2-group $\tilde{\Gamma}$ means that the path integral 
(modulo mutual twistings subtleties as above),
\begin{itemize}
\item sums over $K$ gerbes, and for each $K$ gerbe,
\item sums over ordinary $G$ bundles -- no longer twisted by $K$, as
$BK$ now acts trivially, but with a more subtle restriction on allowed
$G$ bundles, a shadow of the fact that we are gauging a nontrivial
extension of $G$ by
$BK$.
\end{itemize}

These two cases can be subsumed into a more geenral picture
which is most conveniently related by by describing the 2-groups
differently, in terms of what are called crossed modules.
In any event, in this paper we will gauge finite 2-group extensions involving
trivially-acting $BK$, for which the second notion of gauging is a more apt
description.
We will study more general cases in upcoming work.

\subsubsection{Decomposition conjecture}

Consider, as above, gauging a 2-group $\tilde{\Gamma}$ described
formally as an extension of a finite group $G$ by $BK$ for $K$ finite
and abelian,  
\begin{equation}
1 \: \longrightarrow \: BK \: \longrightarrow \: 
\tilde{\Gamma} \: \longrightarrow \: G \: \longrightarrow \: 1.
\end{equation}
This extension determines an element $[\omega] \in H^3(G,K)$.

Because we are gauging a trivially-acting $BK$, one expects that the
theory should possess a global two-form symmetry (distinct from the
quantum symmetry), and so should decompose.

We conjecture that such three-dimensional theories decompose in the form
\begin{equation}  \label{eq:3d-2gp-decomp}
{\rm QFT}\left( [X/\tilde{\Gamma}] \right) \: = \:
{\rm QFT}\left( \coprod_{\rho \in \hat{K}} [X/G]_{\rho(\omega)} \right),
\end{equation}
where $\rho(\omega) \in H^3(G,U(1))$ represents a discrete theta angle,
formally involving a term in the action of the form
\begin{equation}
\int_M \langle \rho, x^* \omega \rangle,
\end{equation}
for $x^* \omega$ as defined in appendix~\ref{sect:homotopy:2-group}.
As we will discuss later, at least on Seifert fibered three-manifolds,
this can be rewritten as
a discrete-torsion-like phase
(of the form discussed in section~\ref{sect:3ddt})
given by the image of $\omega$ under the map
\begin{equation}
H^3(G,K) \: \stackrel{\rho}{\longrightarrow} \: H^3(G,U(1)).
\end{equation}
This is a three-dimensional version of decomposition \cite{Hellerman:2006zs},
whose existence reflects the fact that $[X/\tilde{\Gamma}]$ has
a 2-form symmetry, due to the trivially-acting $BK$.

Next, we will justify this decomposition conjecture by computing 
partition functions for gauged finite 2-groups, and also studying
operator spectra.  In subsequent sections we will check the details in
examples.

\subsubsection{Partition functions}

In this section we will compute partition functions for
$\tilde{\Gamma}$ orbifolds in three dimensions
(for $\tilde{\Gamma}$ a 2-group extension of a finite group $G$ by
a trivially-acting
$BK$).  These are
(weighted) sums over $G$ bundles restricted so that an invariant
vanishes (see appendix~\ref{sect:homotopy:2-group}).
We will see that the resulting partition functions are equivalent to
sums of partition functions of ordinary $G$ orbifolds, weighted by
$C$ field analogues of discrete torsion, 
\begin{equation}
Z\left( [X/\tilde{\Gamma}] \right) \: = \:
\sum_{\rho \in \hat{K}}  Z\left( [X/G]_{\rho(\omega)} \right),
\end{equation}
in accordance with decomposition~(\ref{eq:3d-2gp-decomp}).

In general terms, this is a consequence of the fact, explained
in appendix~\ref{sect:homotopy:2-group}, that $\tilde{\Gamma}$ bundles
on three-manifolds $M$ map to $G$ bundles obeying the constraint
$x^* \omega = 1 \in H^3(M,K)$, 
where $\omega \in H^3(G,K)$ determines the extension
$\tilde{\Gamma}$, and $x: M \rightarrow BG$ determines the $G$ bundle.
Such a constraint is implemented by a projector, proportional to
\begin{equation}
\sum_{\rho \in \hat{K}} \exp\left( \int_M \langle \rho, x^* \omega
 \rangle \right).
\end{equation}
Summing over $\rho \in \hat{K}$ effectively cancels out contributions from
any $G$ bundle for which $x^* \omega \neq 1$.
As we saw for ordinary central extensions in section~\ref{sect:ordinary-case},
inserting such a projection operator in a path integral is equivalent
to working with a sum of theories, one for each $\rho \in \hat{K}$,
each of which is modified by a discrete theta angle defined by
$\rho \in \hat{K}$ and coupling to $x^* \omega \in H^3(M,K)$.  
This gives rise to the present version of
decomposition~(\ref{eq:3d-2gp-decomp}).

At least for Seifert fibered three-manifolds, it is straightforward
to give this construction a much more concrete description, by
describing $x^* \omega$ explicitly in terms of phases derived from the
group cocycle $\omega$.  To do so, we follow the same\footnote{
Our notations differ, but the procedure is identical.  Specifically,
the $\gamma: M \rightarrow BG$ 
used in \cite{Dijkgraaf:1989pz} is the same as $x: M \rightarrow BG$ here,
and the $\alpha \in H^3(G,U(1))$ used there coincides with
$\omega \in H^3(G,K)$ here.  Their analysis is done for $U(1)$ coefficients,
but essentially because $K$ is abelian and in both cases, the group
action on the coefficients is trivial, the argument is otherwise the
same.
} procedure
used in \cite[section 6.5]{Dijkgraaf:1989pz}.  Briefly, given a
triangulation of the three-manifold $M$, 
associate a phase $\omega(g_1, g_2, g_3)$ to
each simplex, and use an ordering to determine whether to multiply
or divide the phase.
(We specialize to Seifert fibered manifolds solely because of potential
practical difficulties in explicitly construction a triangulation.
Given a triangulation, the method of \cite{Dijkgraaf:1989pz} is otherwise
general.)  The result is that $\langle \rho, x^* \omega \rangle$ can be
identified with a discrete-torsion-like phase \cite{Sharpe:2000qt}, 
as described in
section~\ref{sect:3ddt}, for a class in $H^3(G,U(1))$ given by the image of
$\omega \in H^3(G,K)$ under $\rho$, or schematically,
\begin{eqnarray}
H^3(G,K) & \stackrel{\rho}{\longrightarrow} & H^3(G,U(1)) ,
\nonumber\\
\omega & \mapsto & \rho \circ \omega = \rho(\omega).
\end{eqnarray}
We have that on a connected three-manifold $M$,
\begin{equation}
Z_M\left( [X/\tilde{\Gamma}] \right) \: = \:
\sum_{\rho \in \hat{K}} Z_M\left( [X/G]_{\rho(\omega)} \right),
\end{equation}
matching the prediction of decomposition~\ref{eq:3d-2gp-decomp},
with the sum over universes implementing the restriction to
$G$ bundles such that $x^* \omega = 1$.

Next, we specialize to the case of $M = T^3$.  As everything can be
computed explicitly in this case, we will walk through all the
details in order to better explain the idea.

Ordinarily, in a $G$ orbifold on
$T^3$, one would sum over commuting triples $g_1, g_2, g_3 \in G$.
Here, however, because of the 2-group extension, only some triples are
consistent, much as we saw in the case of ordinary central
extensions in section~\ref{sect:ordinary-case}.
As mentioned above, and as described in detail in 
appendix~\ref{sect:homotopy:2-group},
the constraint on $G$ bundles is that $x^* \omega = 1 \in H^3(T^3,K)$.

To understand the result,
we outline here a slightly sloppy computation for the special case of
$T^3$, which will reproduce the $T^3$ result derived rigorously
in appendix~\ref{app:tori}.
To make the 2-group $\tilde{\Gamma}$ more concrete, we imagine associating
$K$-valued wavefunctions $\psi_g$ to $g \in G$, which can then be multiplied by
$K$-valued cocycles, where associativity holds up to
the cocycle $\omega$ as
\begin{equation}
\psi_{g_1 g_2} \psi_{g_3} \: = \: \omega(g_1, g_2, g_3) \, \psi_{g_1} \psi_{g_2
g_3}.
\end{equation}
(Note that adding coboundaries to $\omega$ merely multiplies the products
by phases.)
Then, we can derive a consistency condition on commuting triples,
as follows.
\begin{eqnarray}
\psi_{g_1 g_2} \psi_{g_3} & = & \psi_{g_2 g_1} \psi_{g_3} ,
\nonumber\\
& = & \omega(g_2, g_1, g_3) \psi_{g_2} \psi_{g_1 g_3} ,
\nonumber\\
& = & \omega(g_2, g_1, g_3) \psi_{g_2}\psi_{ g_3 g_1} ,
\nonumber\\
& = &
\frac{ \omega(g_2, g_1, g_3) }{ \omega(g_2, g_3, g_1)}
\psi_{g_2 g_3} \psi_{g_1} ,
\nonumber\\ 
& = & 
\frac{ \omega(g_2, g_1, g_3) }{ \omega(g_2, g_3, g_1)}
\psi_{g_3 g_2} \psi_{g_1} .
\end{eqnarray}
It also equals
\begin{eqnarray}
\psi_{g_1 g_2} \psi_{g_3} & = & \omega(g_1, g_2, g_3) \psi_{g_1} \psi_{g_2 g_3},
\nonumber\\
& = &
 \omega(g_1, g_2, g_3) \psi_{g_1} \psi_{g_3 g_2} ,
\nonumber\\
& = &
\frac{ \omega(g_1, g_2, g_3) }{ \omega(g_1, g_3, g_2) }
\psi_{g_1 g_3} \psi_{g_2} ,
\nonumber\\
& = &
\frac{ \omega(g_1, g_2, g_3) }{ \omega(g_1, g_3, g_2) }
\psi_{g_3 g_1} \psi_{g_2} ,
\nonumber\\
& = &
\frac{ \omega(g_1, g_2, g_3) }{ \omega(g_1, g_3, g_2) }
\omega(g_3, g_1, g_2) 
\psi_{g_3} \psi_{g_1 g_2} ,
\nonumber\\
& = &
\frac{ \omega(g_1, g_2, g_3) }{ \omega(g_1, g_3, g_2) }
\omega(g_3, g_1, g_2) 
\psi_{g_3} \psi_{g_2 g_1} ,
\nonumber\\
& = &
\frac{ \omega(g_1, g_2, g_3) }{ \omega(g_1, g_3, g_2) }
\frac{ \omega(g_3, g_1, g_2) }{ \omega(g_3, g_2, g_1) }
\psi_{g_3 g_2} \psi_{g_1} .
\end{eqnarray}

In order for these two expressions to match, we must require
\begin{equation} \label{eq:3dconstr1}
\frac{ \omega(g_1, g_2, g_3) }{ \omega(g_1, g_3, g_2) }
\frac{ \omega(g_3, g_1, g_2) }{ \omega(g_3, g_2, g_1) }
\frac{ \omega(g_2, g_3, g_1) }{ \omega(g_2, g_1, g_3) }
\: = \: 1
\end{equation}
as an element of $K$, which is the same condition derived
mathematically in appendix~\ref{app:tori}.
(We suspect it may also be possible to use topological defect
lines to give a simple argument, but we leave that for future work.)

We can therefore understand a $\tilde{\Gamma}$ bundle as a
collection of $K$ gerbes and $G$ bundles on $T^3$ defined by
commuting triples $(g_1, g_2, g_3)$ subject to the constraint
\begin{equation}
\epsilon(g_1, g_2, g_3) \: = \: 1
\end{equation}
for
\begin{equation}
\epsilon(g_1, g_2, g_3) \: = \:
\frac{ \omega(g_1, g_2, g_3) }{ \omega(g_1, g_3, g_2) }
\frac{ \omega(g_3, g_1, g_2) }{ \omega(g_3, g_2, g_1) }
\frac{ \omega(g_2, g_3, g_1) }{ \omega(g_2, g_1, g_3) }.
\end{equation}

For the same reasons as discussed for $H^3(G,U(1))$ in
section~\ref{sect:3ddt},
it is straightforward to demonstrate that
\begin{equation}
\epsilon(g_1, g_2, g_3 g_4) \: = \:
\epsilon(g_1, g_2, g_3) \epsilon(g_1, g_2, g_4)
\end{equation}
(and symmetrically), hence using the same argument as in the two-dimensional
case, $\epsilon$ is invariant under simultaneous conjugation\footnote{
We restrict to the same $h$ on each factor because $\epsilon$
is only defined on commuting triples, meaning each pair obeys
$g_i g_j = g_j g_i$.
},
\begin{equation}
\epsilon(h g_1 h^{-1}, h g_2 h^{-1}, h g_3 h^{-1})
\: = \:
\epsilon(g_1, g_2, g_3).
\end{equation}

The partition function of the $\tilde{\Gamma}$ orbifold on $T^3$ then
takes the form\footnote{
The overall factor of $1/|G|$ is standard in orbifolds and ultimately
reflects the fact that the sum is counting bundles with automorphisms,
see e.g.~\cite[equ'n (5.14)]{Freed:1991bn}.
The factors involving $K$ can be found in 
e.g.~\cite[equ'ns (2.31), (2.32)]{Benini:2022hzx}.
} \cite{yujipriv}
\begin{eqnarray}  \label{eq:genl-part-fn}
Z_{T^3}\left( [X/\tilde{\Gamma} ] \right)
& = &
\frac{ | H^0(T^3, K) | }{ | H^1(T^3, K) | } \frac{1}{|H^0(T^3,G)|}
 \sum_{z_1, z_2, z_3 \in K}
{\sum_{g_1, g_2, g_3 \in G}}^{\!\!\!\!\!\prime} \: Z(g_1, g_2, g_3),
\nonumber\\
& = &
\frac{1}{|K|^2 |G|} \sum_{z_1, z_2, z_3 \in K}
{\sum_{g_1, g_2, g_3 \in G}}^{\!\!\!\!\!\prime}\: Z(g_1, g_2, g_3),
\end{eqnarray}
where the prime indicates that the
sum over triples in $G$ is constrained to commuting triples
such that $\epsilon(g_1, g_2, g_3) = 1$.

Now, we can enforce the condition that $\epsilon = 1$ by inserting
a projector
\begin{equation}
\frac{1}{|K|} \sum_{\rho \in \hat{K} } \epsilon_{\rho}(g_1, g_2, g_3)
\end{equation}
where $\epsilon_{\rho}$ is the image of $\epsilon$ under
$\rho: K \rightarrow U(1)$.  
The partition function then has the form
\begin{eqnarray}
Z_{T^3}\left( [X/\tilde{\Gamma} ] \right)
& = &
\frac{1}{|K|^2 |G|} 
|K|^3 \sum_{g_1, g_2, g_3 \in G} 
\frac{1}{|K|} \sum_{\rho \in \hat{K} } \epsilon_{\rho}(g_1, g_2, g_3)
Z(g_1, g_2, g_3),
\nonumber\\
& = &
\sum_{\rho \in \hat{K}}  Z_{T^3}\left( [X/G]_{\epsilon_{\rho})} \right),
\end{eqnarray}
where
\begin{equation}
Z_{T^3}\left( [X/G]_{\epsilon_{\rho})} \right)
\: = \: \frac{1}{|G|} \sum_{g_1, g_2, g_3 \in G} 
 \epsilon_{\rho}(g_1, g_2, g_3)
Z(g_1, g_2, g_3),
\end{equation}
using a standard normalization (compare e.g.~\cite[equ'n (5.14)]{Freed:1991bn}).
Each factor $\epsilon_{\rho}$ is precisely a $C$ field analogue of
discrete torsion, as reviewed in section~\ref{sect:3ddt},
and coincides with the quantity we earlier labelled $\rho(\omega)$.

Thus, we see that for the special case of $T^3$, partition functions
are consistent with the decomposition conjecture~\ref{eq:3d-2gp-decomp}.
As outlined at the beginning, the same argument applies for any
three-manifold.  The only real difference on other three-manifolds
is that there may be dilaton-type Euler counterterm shifts,
as discussed
in e.g.~\cite{Hellerman:2006zs}, which vanish on $T^3$ as 
$\chi(T^3) = 0$.  Modulo such trivial
counterterms,  
on any connected three-manifold,
\begin{equation}
Z\left( [X/\tilde{\Gamma}] \right) \: = \:
\sum_{\rho \in \hat{K}}  Z\left( [X/G]_{\epsilon_{\rho}} \right).
\end{equation}
This is precisely the statement of decomposition~(\ref{eq:3d-2gp-decomp}),
at the level of partition functions.

To summarize, 
we see that
inserting a projection operator to enforce the constraint on $G$-twisted
sectors makes manifest
the statement that the partition function of the 2-group orbifolds
equals
the partition function for
a sum of three-dimensional orbifolds, each twisted by an
$\epsilon_{\rho}$ which is \cite{Sharpe:2000qt} a three-dimensional analogue of
discrete torsion.  In this fashion, we recover 
decomposition~(\ref{eq:3d-2gp-decomp}), at the level of partition functions,
in close analogy with the description in section~\ref{sect:ordinary-case} of
decomposition in two-dimensional orbifolds.

As an aside, previously in two-dimensional theories with a one-form
symmetry given by a trivially-acting $K$, we saw universes
enumerated by irreducible representations of $K$,
see e.g.~\cite{Hellerman:2006zs}.
Here, since we have a 2-form symmetry and trivially-acting $BK$,
one might have naively guessed that universes would be enumerated
by representations of $BK$, at variance with the
conjecture above.  However, we examine decomposition for both
1-form and 2-form symmetries formally in appendix~\ref{app:duality},
and observe there that in both cases, universes appear to be enumerated
by representations of $K$, so the form of the conjecture above is consistent.

\subsubsection{Local operators}

So far we have given a general justification of the decomposition conjecture
for gauged 2-groups using partition functions.  Let us briefly outline an
analogous argument using local operators.  In two dimensional orbifolds
with trivially-acting subgroups, the twist fields associated to trivially-acting
group elements form dimension-zero operators, and the projectors (onto
universes) are constructed from linear combinations of those projectors.
In three dimensions, 
when gauging a one-form symmetry, from the general theory of
topological defect lines, the theory contains monopole operators, which
play an analogous role.  Briefly, the monopole operators are endpoints
of real codimension two lines corresponding to the gauged one-form symmetry,
just as gauging an ordinary (zero-form) symmetry results in real
codimension one walls.  Two-spheres surrounding the monopole operators have
$K$ gerbes, just as circles surrounding two-dimensional twist fields
carry bundles.

In any event, given a trivially-acting gauged $BK$ symmetry,
the resulting three-dimensional theory will contain monopole operators,
which are closely analogous to two-dimensional twist fields, and can be
used to build projectors.

For example, in a gauged $B {\mathbb Z}_k$, the monopole operators will
generate ${\mathbb Z}_k$ gerbes on $S^2$, which are classified by
$H^2(S^2,{\mathbb Z}_k) = {\mathbb Z}_k$.  As those gerbes on $S^2$
are all generated by powers of one gerbe, there will be one monopole operator
which generates the others, call it $\hat{z}$, and which obeys
$\hat{z}^k = 1$.  Given such operators, one can build projectors, as linear
combinations of the form
\begin{equation}
\Pi_m \: = \: \frac{1}{k} \sum_{j=0}^{k-1} \xi^{jm} \hat{z}^j,
\end{equation}
for $\xi = \exp(2 \pi i/k)$, which from $\hat{z}^k = 1$ are easily checked
to obey
\begin{equation}
\Pi_m \Pi_n \: = \: \Pi_m \delta_{m,n}, \: \: \:
\sum_{m=0}^{k-1} \Pi_m \: = \: 1.
\end{equation}

\subsection{Example:  $G = 1$, $K = {\mathbb Z}_2$}

Let us consider the orbifold $[X/B {\mathbb Z}_2]$ for a moment,
where the $B {\mathbb Z}_2$ acts trivially, in the sense that all line
operators in the theory are invariant under the $B {\mathbb Z}_2$.

Then, at a path integral level, the orbifold $[X/B {\mathbb Z}_2]$ involves
a sum over ${\mathbb Z}_2$ gerbes, but each of the gerbe sectors is identical,
much as in a two-dimensional orbifold by a group that acts completely trivially.

At the level of operators, gauging the $B {\mathbb Z}_2$ results in 
monopole operators, which generate ${\mathbb Z}_2$ gerbes on spheres
surrounding the operators, much as twist fields generate branch cuts and
hence
bundles on surrounding circles in two-dimensional theories.

Since the $B {\mathbb Z}_2$ acts trivially, the monopole operators
commute with all local operators present in the original theory,
we see that the full set of operators in the gauged theory is just two
copies of the operators of the original theory.  Furthermore, since the
monopole operators generate ${\mathbb Z}_2$ gerbes on surrounding $S^2$'s,
and the product of a nontrivial ${\mathbb Z}_2$ gerbe with itself is trivial,
we see that if $\hat{z}$ denotes a monopole operator, then $\hat{z}^2 = 1$,
and so we can build projection operators
\begin{equation}
\Pi_{\pm} \: = \: \frac{1}{2} \left( 1 \pm \hat{z} \right),
\end{equation}
which implement a decomposition.

In particular, in these circumstances,
\begin{equation}
[X / B {\mathbb Z}_2 ] \: = \: X \, \coprod \, X,
\end{equation}
as expected from decomposition~(\ref{eq:3d-2gp-decomp}).
(Here, we use the fact that $\rho(\omega) = 1$ for all $\rho \in \hat{K}$,
as $\omega$ itself is trivial.)

\subsection{Example:  $G = {\mathbb Z}_2 = K$}

Let us begin with a very simple example.
Consider the case of a two-group extension of the form
\begin{equation}
1 \: \longrightarrow \: B {\mathbb Z}_2 \: \longrightarrow \:
\tilde{\Gamma} \: \longrightarrow \: {\mathbb Z}_2 \: \longrightarrow \: 1.
\end{equation}
As discussed in appendix~\ref{app:z2}, $H^3({\mathbb Z}_2, {\mathbb Z}_2) = 
{\mathbb Z}_2$, so there is a nontrivial 2-group extension $\tilde{\Gamma}$
of this form.

In this case, it is straightforward to check that
$\epsilon(g_1, g_2, g_3)$ is the identity in ${\mathbb Z}_2$ for all
triples $g_{1-3} \in {\mathbb Z}_2$,
so there is no additional constraint on $G$ bundles on $T^3$
(beyond pairwise commutivity) to lift to a $\tilde{\Gamma}$ bundle.

It is then straightforward to compute the $T^3$ partition function
from~(\ref{eq:genl-part-fn}), yielding
\begin{eqnarray}
Z_{T^3}\left( [X/\tilde{\Gamma} ] \right)
& = &
\frac{1}{|K|^2 |G|} \sum_{z_1, z_2, z_3 \in K}
\sum_{g_1, g_2, g_3 \in G} Z(g_1, g_2, g_3),
\nonumber\\
& = &
\frac{|K|}{|G|} \sum_{g_1, g_2, g_3 \in G} Z(g_1, g_2, g_3),
\nonumber\\
& = & Z_{T^3} \left( \coprod_{\hat{K}} [X/G] \right),
\end{eqnarray}
as expected from decomposition~(\ref{eq:3d-2gp-decomp}).

In this case, $\rho(\omega) = 1$ for all $\rho \in \hat{K}$.
Although $H^3(G,U(1)) = {\mathbb Z}_2$ for $G = {\mathbb Z}_2$,
the group $G = {\mathbb Z}_2$ is in some sense too small to have
any nontrivial phases resulting from analogues of discrete torsion.

The reader should also note that we get this decomposition for
both 2-group extensions $\tilde{\Gamma}$ indexed by
$H^2({\mathbb Z}_2,{\mathbb Z}_2) = {\mathbb Z}_2$,
implying that they are physically equivalent to one another.
(Analogous relations were seen in decomposition of two-dimensional
theories with one-form symmetries in \cite{Hellerman:2006zs}, in which
different gerbes are described by the same physical theory.)

\subsection{Example:  $G = ({\mathbb Z}_2)^3$, $K = {\mathbb Z}_2$}

Write $G = ({\mathbb Z}_2)^3 = \langle a, b, c \rangle$.
Let us pick an extension of $G$ by $BK$ corresponding to the
element of $H^3(G,K)$ given by $(-)^{a_1 b_2 c_3}$ in appendix~\ref{app:z23}.

Then, the commuting triples $g_{1-3}$ for which $\epsilon(g_1, g_2, g_3) \neq
1 \in K$ include, for example,
$(ax,by,cz)$ and their permutations, where
\begin{equation} \label{eq:ex:omitted}
x \in \{1, b, c, bc\}, \: \: \:
y \in \{1, a, c, ac\}, \: \: \:
z \in \{1, a, b, ab\}.
\end{equation}

The partition function of $[X/\tilde{\Gamma}]$ then has the form
\begin{equation}
Z_{T^3}\left( [X/\tilde{\Gamma}] \right) \: = \:
\frac{1}{|K|^2 |G|} \sum_{z_{1-3}\in K} {\sum_{g_{1-3}\in G}}^{\!\!\!\prime} \: Z(g_1, g_2, g_3)
\: = \:
\frac{|K|}{|G|} {\sum_{g_{1-3}\in G}}^{\!\!\!\prime} \: Z(g_1, g_2, g_3),
\end{equation}
where the prime indicates that some of the $G$-twisted sectors are omitted.

For the trivial representation $1 \in \hat{K}$, $\epsilon_1(g_1, g_2, g_3) = 1$,
but for the nontrivial representation $1 \in \hat{K}$,
$\epsilon_{\rho}(g_1, g_2, g_3)$ corresponds to the discrete-torsion-like
phase~(\ref{eq:dw-phases}) corresponding to the cocycle $\omega_4
\in H^3(G,U(1))$ listed in appendix~\ref{app:z23}, essentially because
the $\omega_4$ cocycle has the same form as the chosen element
of $H^3(G,K)$ above:  $\omega_4(g_1, g_2, g_3) = (-)^{a_1 b_2 c_3}$ also.
That discrete-torsion-like phase equals $-1$ on precisely the
triples that are omitted from the $[X/\tilde{\Gamma}]$ orbifold,
namely sectors of the form $(ax, by, cz)$ and their permutations,
for $x$, $y$, $z$ as in~(\ref{eq:ex:omitted}).
Sectors that are not omitted
include $(g,g,g)$ for $g$ any element of $( {\mathbb Z}_2 )^3$.

Putting this together, we see
\begin{equation}
Z_{T^3}\left( [X/\tilde{\Gamma}] \right) \: = \:
Z_{T^3}\left( [X/G] \, \coprod \,
[X/G]_{\omega_4} \right),
\end{equation}
matching the prediction of decomposition~(\ref{eq:3d-2gp-decomp}) for this case.
The sectors that are omitted in the $\tilde{\Gamma}$ orbifold cancel out
between the two $G$ orbifolds, realizing a `multiverse interference
effect' as usual.

\subsection{Example: $G = ({\mathbb Z}_2)^2 = K$}

In this case, it is straightforward to check that the
discrete-torsion-like phase factors $\omega(\rho)$ are all trivial
for any extension class in $H^3(G,K)$ and any $\rho \in \hat{K}$,
hence in this case our conjecture~(\ref{eq:3d-2gp-decomp}) predicts
\begin{equation}
{\rm QFT}\left( [X/\tilde{\Gamma}] \right) \: = \:
{\rm QFT}\left( \coprod_{\rho \in \hat{K}} [X/G] \right).
\end{equation}

We can check this by computing the $T^3$ partition function.
In this case, for $G=K=({\mathbb Z}_2)^2$, it is straightforward
to check that $\epsilon = 1$ holds automatically for every
$[\omega] \in H^3( G, K )$, so there is no constraint on
commuting triples $(g_1, g_2, g_3)$.
Then,
from the general formula~(\ref{eq:genl-part-fn}),
\begin{eqnarray}
Z_{T^3}\left( [X/\tilde{\Gamma} ] \right)
& = &
\frac{1}{|K|^2 |G|} \sum_{z_1, z_2, z_3 \in K}
\sum_{g_1, g_2, g_3 \in G} Z(g_1, g_2, g_3)
,
\nonumber\\
& = &
\frac{ |K| }{ |G| } \sum_{g_1, g_2, g_3 \in G} Z(g_1, g_2, g_3)
,
\nonumber\\
& = &
|K| Z_{T^3}\left( [X/G] \right),
\end{eqnarray}
which is consistent with the prediction of decomposition.

\section{Interpretation: sigma models on 2-gerbes}
\label{sect:sigma-2gerbe}

These orbifolds by 2-groups have a more formal description as
realizations of sigma models on 2-gerbes, closely analogous to
sigma models on gerbes as described in
\cite{Pantev:2005rh,Pantev:2005zs,Pantev:2005wj}.

Briefly, gerbes are closely analogous to principal bundles.
A $n$-($G$-)gerbe is essentially a fiber bundle whose fibers
are `groups' $B^n G$ of higher-form symmetries.  As a result,
a sensibly-defined sigma model with target such a gerbe should
admit a global $B^n G$ symmetry, corresponding to translations along the
fibers of the gerbe.

Because  the `group' $BG =
[{\rm point}/G]$, a $G$-gerbe -- a fiber bundle with fiber $BG$ --
can be locally
presented as a quotient in which a subgroup acts trivially.
This was utilized in the previous work
\cite{Pantev:2005rh,Pantev:2005zs,Pantev:2005wj}
to construct sigma models on gerbes, presented as
orbifolds and gauge theories
with trivially-acting subgroups.

Now, this glosses over a number of subtleties, including questions about
non-uniqueness of presentations (dealt with by identifying a sigma model
on a stack or gerbe with a universality class of RG flow), potential
modular invariance and unitarity issues in orbifolds, seeming moduli
mismatches, and most important for decomposition, violations of
the cluster decomposition axiom, which were discussed in
\cite{Pantev:2005rh,Pantev:2005zs,Pantev:2005wj,Hellerman:2006zs}.

In any event, from the same reasoning,
orbifolds by 2-groups with trivially-acting one-form symmetries
appear to be presentations of
sigma models on 2-gerbes, just as sigma models on ordinary gerbes
are realized in terms of gauge theories with trivially-acting
(ordinary) subgroups \cite{Pantev:2005rh,Pantev:2005zs,Pantev:2005wj}.

As discussed in \cite[section 2]{Hellerman:2010fv},
a map $f: Y \rightarrow {\cal G}$, for ${\cal G}$ a (banded) $G$-gerbe over $M$
($G$ assumed finite),
defines\footnote{
In fact, the map $f$ is equivalent to the map $\tilde{f}$ plus a specific
choice of trivialization of $\tilde{f}^* {\cal G}$.
} a map $\tilde{f}: Y \rightarrow M$ with a trivialization of
$\tilde{f}^*
{\cal G}$.
If $\dim Y = 2$, this gives a restriction on the degree of $\tilde{f}$.
Explicitly, let $\pi: {\cal G} \rightarrow M$ be projection,
then $\tilde{f} = \pi \circ f$, and $\tilde{f}^* {\cal G}$ has a canonical
trivialization.  This trivialization may be clearer to the reader in the closely
related case of bundles.  Given a map $g: Y \rightarrow E$ for some bundle
$\pi: E \rightarrow M$, we can define $\tilde{g} = \pi^* g$,
and then as
\begin{equation}
\tilde{g}^* E \: = \: \{(y,e) \in Y \times E \, | \, \tilde{g}(y) = \pi(e) \},
\end{equation}
there is a trivialization $Y \rightarrow \tilde{g}^* E$ given by
$y \mapsto (y,\tilde{g}(y))$.  The same analysis applies to gerbes.

So, we have that a map $f: Y \rightarrow {\cal G}$ defines a map
$\tilde{f}: Y \rightarrow M$ such that $\tilde{f}^* {\cal G}$ is
trivializable.  As discussed in \cite[section 2]{Hellerman:2010fv},
if $\dim Y = 2$, this implies a restriction on degrees.  If
the characteristic class of ${\cal G}$ is $\omega \in H^2( M, G)$ ($G$ finite),
then $\tilde{f}^* \omega = 0 \in H^2(Y, G)$.  For example,
if $Y = {\mathbb P}^1$ and $M = {\mathbb P}^N$, with
$\tilde{f}: {\mathbb P}^1 \rightarrow {\mathbb P}^N$ of degree $d$,
and $G = {\mathbb Z}_k$, then $\tilde{f}^* \omega = d \omega$,
and $d \omega = 0 \in H^2( {\mathbb P}^N, {\mathbb Z}_k)$ means
$d \omega \equiv 0 \mod k$, that the product of $d$ and the
characteristic class is divisible by $k$.

If the dimension of $Y$ is not two, then one still has a constraint
that $\tilde{f}^* {\cal G}$ is trivializable, which does restrict the
possible maps $\tilde{f}$; however, that restriction will not be describable
as simply as a restriction on map degrees.

Briefly, the same formal arguments apply to (banded analogues of)
2-gerbes.  Just as for ordinary gerbes, a map $f: Y \rightarrow {\cal G}$,
for ${\cal G}$ a 2-($G$-)gerbe over $M$, from essentially the same
argument as before,
one gets a map $\tilde{f}: Y \rightarrow M$ with a restriction on
degrees, following from the statement that $\tilde{f}^* {\cal G}$ is
trivializable (and so has vanishing characteristic class in
$H^3(Y,G)$).

\section{Analogues in other dimensions and other degrees}
\label{sect:higher}

\subsection{Decomposition in higher-dimensional orbifolds}

In this section, we make some conjectures for how this program could
be continued into higher dimensions, by observing that the arguments
we have applied to ordinary central extensions and 2-group extensions
also apply, with only minor modifications, to higher-group extensions.

Consider orbifolds in $d$ dimensions.  Specifically,
consider gauging a higher-group extension
\begin{equation}
1 \: \longrightarrow \: B^{d-2} K \: \longrightarrow \: \tilde{\Gamma} \:
\longrightarrow \: G \: \longrightarrow \: 1,
\end{equation}
for $K$ a finite abelian group, classified by an element
$[\omega] \in H^{d}(G,K)$.  The orbifold $[X/\tilde{\Gamma}]$ has the
structure of a $[X/G]$ orbifold but with a restriction on the $G$ sectors,
namely that they trivialize a coboundary-invariant constructed from
$\omega$, or explicitly $x^* \omega = 1$ in the notation of
appendix~\ref{app:homotopy}.  
For example, on $T^d$, we require that commuting $d$-tuples
$g_1, \cdots, g_d$ also obey
\begin{equation}  \label{eq:dtuple:restr}
\epsilon(g_1, \cdots, g_d) \: = \: 1 \: \in \: K,
\end{equation}
for
\begin{equation}
\epsilon(g_1, \cdots, g_d) \: = \: \prod_{{\rm perm's} \: \sigma}
 \omega(g_{\sigma(1)},
\cdots, g_{\sigma(d)} )^{{\rm sgn}\, \sigma},
\end{equation}
as outlined in appendix~\ref{app:tori}.

The reader should note in passing that the phase $\epsilon$ above, for 
coefficients in any abelian group, obeys
standard properties of discrete-torsion-like phases, specifically,
\begin{itemize}
\item the phase $\epsilon$ is invariant under coboundaries, 
and so is well-defined
on cohomology $H^d(G,U(1))$,
\item the phase $\epsilon$ is a homomorphism in the sense that
\begin{equation}
\epsilon(ab,g_3,\cdots,g_{d+1}) \: = \: \epsilon(a,g_3, \cdots, g_{d+1})
\, \epsilon(b,g_3,\cdots,g_{d+1}),
\end{equation}
(and similarly for products in other positions, 
from the antisymmetry of $\epsilon$),
as can be verified from the identity
\begin{equation}
{\prod_{ {\rm perm's}\: \sigma}}^{\!\!\!\prime} \:
(d\omega)\left(g_{\sigma(1)}, \cdots, g_{\sigma(d+1)} \right)^{
{\rm sgn}\:\sigma} \: = \: 1,
\end{equation}
for permutations of the $(d+1)$-tuple $(a, b, g_3, \cdots, g_{d+1})$,
where the prime indicates that we restrict to permutations preserving
the order of $a$, $b$,
\item the phase $\epsilon(g_1,\cdots,g_d)$ 
is invariant under $SL(n,{\mathbb Z})$ actions
on the group elements, as is straightforward to verify from the homomorphism
property.
\end{itemize}

Returning to partition functions,
the restriction above on $G$ bundles can be implemented by inserting
a projector, which (as discussed previously) is equivalent to a decomposition
into universes $[X/G]$ weighted by a discrete theta angle coupling to
$x^* \omega$, in the notation of appendix~\ref{app:homotopy}.

In the special case of $T^d$,
the restriction above to $d$-tuples obeying~(\ref{eq:dtuple:restr})
is equivalent to inserting a projection operator
in an ordinary $[X/G]$ orbifold, with projector which on $T^d$ takes the
form
\begin{equation}
\frac{1}{|K|} \sum_{\rho \in \hat{K}} \epsilon_{\rho}(g_1, \cdots, g_d),
\end{equation}
where $\epsilon_{\rho} \in U(1)$ is the image of $\epsilon$ under
$\rho: K \rightarrow U(1)$.
The resulting $T^d$ partition function is the same as that of
a sum of partition functions of $[X/G]$ orbifolds, each with
a discrete-torsion-like phase factor defined by
$\epsilon_{\rho}$.

Thus, in higher dimensions, based on the partition function
analysis above, we expect that the
$[X/\tilde{\Gamma}]$ orbifold decomposes:
\begin{equation}
{\rm QFT}\left( [X/\tilde{\Gamma}] \right) \: = \:
{\rm QFT}\left( \coprod_{\rho \in \hat{K}} [X/G]_{\rho(\omega)} \right),
\end{equation}
(for $\rho(\omega)$ indicating a discrete theta angle $\rho$ coupled to
$x^* \omega$,)
which at least in special cases can be expressed in the form
\begin{equation}
{\rm QFT}\left( [X/\tilde{\Gamma}] \right) \: = \:
{\rm QFT}\left( \coprod_{\rho \in \hat{K}} [X/G]_{\rho(C)} \right),
\end{equation}
for $\rho(C)$ expressing elements of higher-dimensional analogues of
discrete torsion.
(Interpreted literally as a sigma model, this theory should only
be understood as a low-energy effective action, of course, though 
this should also be a prototype for theories in $d$ dimensions.)

It is also straightforward to outline the origin of projectors in this
language.  In two dimensional orbifolds, the projectors onto the universes
are constructed as linear combinations of the twist fields associated to
trivially-acting group elements.  Now, in a $d$ dimensional theory, if
we gauge a $p$-form symmetry, then in the language of topological
defect lines (see e.g.~\cite{Chang:2018iay}),
one gets a real codimension $(p+1)$ object
that generalizes the branch cuts of an orbifold, and which terminates on
a real codimension $(p+2)$ object, which is the analogue of a twist field.

So, work in $d$ dimensions, and gauge a (trivially-acting) $(d-2)$-form
symmetry.  In principle, this should result in a theory with a global
$(d-1)$-form symmetry, and hence a decomposition.
Because we have gauged a $(d-2)$-form symmetry, we get a real codimension
$(d-1)$ object, an analogue of the two-dimensional branch cut,
which terminates at a real codimension $d$ object (an analogue of a twist
field), which in $d$ dimensions is pointlike.  Those pointlike objects,
those analogues of twist fields, could then be used to construct
projectors.

\subsection{Interpretation: higher-dimensional sigma models}
\label{sect:higher-sigma}

In this paper we have discussed how maps from 2-manifolds
into ordinary gerbes
and maps from 3-manifolds into 2-gerbes define maps into spaces with
restrictions on degrees (following from the constraint that the pullback
of the gerbe be trivial).

There is a very closely analogous story for higher gerbes,
which we outline in this section (slightly generalizing
\cite[section 2]{Hellerman:2010fv}).
Maps into ($m$-)$G$-gerbes are closely related to maps into underlying spaces
with restrictions on degrees.  Consider a map $f$ from a space $Y$
into a ($m$-)gerbe ${\cal G} \rightarrow M$.  Composing with the
projection gives a map $\tilde{f}: Y \rightarrow M$.
The map $f$ defines a section
of $\tilde{f}^* {\cal G}$, almost by definition, hence it trivializes
$\tilde{f}^* 
{\cal G}$.

As a consequence, the map $\tilde{f}$ induces
\begin{equation}
\tilde{f}^*: \: H^{m+1}(M, G) \: \longrightarrow \: H^{m+1}(Y, G).
\end{equation}
The characteristic class of the $m$-gerbe ${\cal G}$ must be in the kernel
of that map, hence there is a restriction on possible maps $\tilde{f}$.

In particular, a map $f: Y \rightarrow {\cal G}$ is equivalent to a map
$\tilde{f}: Y \rightarrow M$,
trivializing the characteristic class of the gerbe,
together with a specific choice of trivialization of the $m$-gerbe
$\tilde{f}^* {\cal G}$, which is an $(m-1)$-gerbe over $B$.

Depending upon the circumstances, this may imply a restriction on the
map $\tilde{f}$.  For example,
if ${\cal G}$ is an $m$-gerbe and $\dim Y \leq m$, the map $\tilde{f}$ is
unconstrained, since the pullback of the characteristic class is
an element of $H^{m+1}(Y,G) = 0$, so all maps are in the kernel.

On the other hand, suppose we have an $m$-gerbe and
$\dim Y > m$.  (For example, a four-dimensional low-energy effective
sigma model mapping into a 1-gerbe, 2-gerbe, or 3-gerbe.) 
In this case, the map $\tilde{f}$ is constrained, but depending upon
the relative values of $m$ and $\dim Y$, the restriction may be on e.g.~lower
homotopy.

\section{Analogues in Chern-Simons theories in three dimensions}
\label{sect:cs}

It is well-known that gauging the $B {\mathbb Z}_2$ central symmetry
of $SU(2)$ Chern-Simons theory in three dimensions results in an $SO(3)$
Chern-Simons theory.  Briefly, the path integral sums over
${\mathbb Z}_2$ gerbes and gerbe-twisted $SU(2)$ bundles with connection,
for which
bundle transition functions only close up to gerbe transition functions
on triple overlaps; the resulting path integral is precisely a path
integral over $SO(3)$ bundles with connection, for which the second
Stiefel-Whitney class $w_2$ coincides with the gerbe characteristic
class, and the third Stiefel-Whitney class is determined by a Steenrod
square as $w_3 = {\rm Sq}^1(w_2)$.

In that case, the $B {\mathbb Z}_2$ acted nontrivially on line operators,
specifically as phases determined by the $n$-ality of the representation
(partially) defining the Wilson line.

We could consider more general situations, in which the one-form symmetry
group maps to an action on the center, but with a nonzero kernel.
In general, consider a 2-group $\Gamma$ defined by a crossed module
$\{d: A \rightarrow H\}$, where $A$ is abelian and the image of $d$
is contained within the center of the group $H$.  If we let
$K$ denote the kernel of $d$, and $G = H / {\rm im}\, A$, then
\begin{equation}
1 \: \longrightarrow \: K \: \longrightarrow \: A \: \longrightarrow \:
H \: \longrightarrow \: G \: \longrightarrow \: 1,
\end{equation}
which defines an element $\omega \in H^3(G,K)$.  In principle, if $G$
is, for example, a Lie group, but we are only concerned with flat
bundles, then the same homotopy computations of
appendix~\ref{sect:homotopy:2-group} imply that (flat) $\Gamma$ bundles
map to (flat) $G$ bundles obeying the constraint that $\phi^* \omega = 0$.

Such a constraint can be implemented via a decomposition, and flat bundles
arise in Chern-Simons theories, so we have a prediction:
\begin{equation}
\mbox{Chern-Simons}(H) / BA \: = \: \coprod_{\theta \in \hat{K}}
\mbox{Chern-Simons}(G)_{\theta},
\end{equation}
where the $\theta$ are discrete theta angles coupling to $\phi^* \omega$,
and for levels such that the Chern-Simons theories are defined.

For example, consider an $SU(2)$ Chern-Simons with an action of
$B {\mathbb Z}_4$, which maps to the central one-form symmetry of $SU(2)$,
with a $B {\mathbb Z}_2$ kernel which leaves all line operators invariant.
In this case, we predict
\begin{equation}
\mbox{Chern-Simons}(SU(2)) / B {\mathbb Z}_4 \: = \:
\mbox{Chern-Simons}(SO(3))_+ \: \coprod \:
\mbox{Chern-Simons}(SO(3))_-.
\end{equation}

This form of decomposition will be discussed in detail in upcoming work.

\section{Conclusions}

In this paper we have discussed 2-group orbifolds and their decomposition.
Because these theories involve the gauging of a trivially-acting
one-form symmetry, they possess a global two-form symmetry, implying
a decomposition.
The pattern followed is very similar to two dimensions:  the
twisted sectors of the 2-group orbifolds look like twisted sectors
of ordinary orbifolds obeying a constraint, and that constraint is 
implemented by the decomposition.

In our analysis, we specialized to 2-groups that were analogues of
central extensions, defined in part by trivial group actions of $G$
on $K$.  It would be interesting to consider more general cases;
such analyses are left for future work.

One direction that would be interesting to pursue would be to
deform 2-group orbifolds by turning on $C$ field flux,
in the same way that one can turn on discrete torsion to deform
ordinary two-dimensional orbifolds.  Decomposition in orbifolds with
discrete torsion was discussed in \cite{Robbins:2020msp}.
A related direction that would be interesting to pursue would be
analogues of quantum symmetries in 2-group orbifolds,
generalizing the results of
\cite{Robbins:2021ibx}.

\section*{Acknowledgements}

We would like to thank S.~Gukov, U~Schreiber, Y.~Tachikawa, B.~T\"oen,
and M.~Yu for useful discussions.
T.P. was partially supported by NSF/BSF grant DMS-2200914, 
NSF grant DMS-1901876, and Simons Collaboration grant number 347070.
D.R. and T.V. were partially supported by
NSF grant PHY-1820867.
E.S. was partially supported by NSF grant
PHY-2014086.

\appendix

\section{Homotopy theory}
\label{app:homotopy}

In this section we will give more rigorous justifications 
of statements appearing in the main text
that various $\Gamma$ bundles project to $G$ bundles
obeying a restriction,
utilizing homotopy theory.  We first describe such restrictions in the
case of ordinary central extensions, as a warm-up exercise for the
reader, then turn to 2-group extensions.

\subsection{Classification of bundles for central extensions}
\label{app:class:bundles}

Let $\omega \in H^2(G,K)$, for $G$ and $K$ finite groups and $K$ abelian,
and associate a central extension $\Gamma$:
\begin{equation}  \label{eq:ses-basic}
1 \: \longrightarrow \: K \: \longrightarrow \: \Gamma \: \longrightarrow \:
G \: \longrightarrow \: 1.
\end{equation}
We want to understand $\Gamma$ bundles, which is to say,
${\rm Map}(M, B \Gamma)$.

Now, from the surjective map in the central extension, there is a map
$B \Gamma \rightarrow BG$.

Furthermore, since $\Gamma$ is a central extension of $G$ by $K$,
$\Gamma \rightarrow G$ is a principal $K$ bundle, hence classified by
a map $G \rightarrow BK$, meaning that $\Gamma$ is the fiber product
\begin{equation}
\xymatrix{
\Gamma \ar[r] \ar[d] & {\rm point} \ar[d] \\
G \ar[r] & BK
}
\end{equation}
Taking $B$, we get the fiber product
\begin{equation}
\xymatrix{
B \Gamma \ar[r] \ar[d] & {\rm point} \ar[d] \\
BG \ar[r]^-{\omega} & K(K,2)
}
\end{equation}
Put another way, if we apply the functor $B(-)$ to the
short exact sequence~(\ref{eq:ses-basic}) we get that $B \Gamma \rightarrow BG$
is a $BK$ principal bundle, and so classified by a map
$BG \rightarrow B^2 K = K(K,2)$, which gives the diagram above.

In passing, if $K$ is abelian but not central,
then the $G$ action on $K$ gives a twisted form of $K$,
a sheaf of groups ${\cal K} \rightarrow BG$ which is locally isomorphic
to $K \times BG$.  The extension class $\omega$ is then an element of
$H^2(BG,{\cal K})$ (instead of $H^2(BG,K)$),
equivalently a section of $\Gamma(BG,K({\cal K},2))$.
In this case, $B\Gamma$ is the homotopy intersection of this section and
the zero section of $K({\cal K},2)$.

Returning to central extensions,
one has
the diagram
\begin{equation}
\xymatrix{
{\rm Map}(M,B\Gamma) \ar[r] \ar[d] 
& {\rm Map}(M, {\rm point}) \: = \: {\rm point}
\ar[d]
\\
{\rm Map}(M,BG) \ar[r] & {\rm Map}(M,K(K,2)) \: = \: H^2(M,K),
}
\end{equation}
where the bottom map sends $x \in {\rm Map}(M,BG) \: \mapsto \:
x^* \omega$.
An element $x \in {\rm Map}(M,BG)$
will be in the image of an element of ${\rm Map}(M,B \Gamma)$
precisely when $x^* \omega = 1 \in H^2(M,K)$.  
We will examine the implications of this on tori in
section~\ref{app:tori}.

It may be helpful to observe that $x^* \omega$ can also be understood
as the image of the isomorphism class of the $G$ bundle in
$H^1(M,G)$ under the Bockstein homomorphism
\begin{equation}
H^1(M,G) \: \longrightarrow \: H^2(M,K).
\end{equation}
If we momentarily drop the assumption that $G$ be finite, then for
$G = SO(n)$, $\Gamma = {\rm Spin}(n)$, the quantity we label
$x^* \omega$ would coincide with the second Stiefel-Whitney class of
the $G$ bundle.

\subsection{Classification of 2-group bundles}
\label{sect:homotopy:2-group}

Now, let us repeat that analysis for bundles of 2-groups constructed
analogously as extensions.

Let $\omega \in H^3(G,K) = H^3(BG,K)$.
Associated to this is a 2-group, which we describe as a
crossed module $\Gamma_{\bullet}$,
\begin{equation}
\Gamma_{\bullet} \: = \: \{ \Gamma_1 \: \stackrel{d}{\longrightarrow} \:
\Gamma_0 \},
\end{equation}
which sits in the sequence
\begin{equation}
1 \: \longrightarrow \: K \: \longrightarrow \: \Gamma_1 \:
\stackrel{d}{\longrightarrow} \: \Gamma_0 \: \longrightarrow \: G \:
\longrightarrow \: 1.
\end{equation}
The classifying 2-stack $B \Gamma_{\cdot}$ is connected, with homotopy
groups
\begin{equation}
\pi_0(B \Gamma_{\bullet} ) \: = \: 0, \: \: \:
\pi_1(B \Gamma_{\bullet} ) \: = \: G, \: \: \:
\pi_2(B \Gamma_{\bullet} ) \: = \: K.
\end{equation}

Let $M$ be a compact oriented manifold.
We want to understand Map$(M, B \Gamma_{\bullet})$.

To that end, note that there is a map
\begin{equation}
B \Gamma_{\bullet} \: \longrightarrow \: BG,
\end{equation}
arising as the first stage of the Postnikov tower of $B \Gamma_{\bullet}$,
and in fact $B \Gamma_{\bullet}$ is a fiber square
\begin{equation}
\xymatrix{
B \Gamma_{\bullet} \ar[r] \ar[d] &  {\rm point}
 \ar[d] \\
BG \ar[r]^-{\omega} & K(K,3),
}
\end{equation}
(from the definition of $B \Gamma_{\bullet}$ as a homotopy type).
In the bottom map, we interpret $\omega \in H^3(G,K)$ by writing
\begin{equation}
H^3(G,K) \: = \: H^3(BG,K) \: = \: {\rm Map}\left(BG, K(K,3) \right).
\end{equation}

From the fiber square above, we derive the square
\begin{equation}
\xymatrix{
{\rm Map}(M, B\Gamma_{\bullet}) \ar[r] \ar[d] & 
{\rm Map}(M,{\rm point}) \: = \: {\rm point}
\ar[d] \\
{\rm Map}(M,BG) \ar[r] & {\rm Map}(M,K(K,3)) \: = \: H^3(M,K).
}
\end{equation}
which constrains possible maps (hence possible $G$ bundles on $M$).

Note that since
\begin{equation}
\omega \: \in \: H^3(G,K) \: = \: H^3(BG,K) \: = \:
{\rm Map}\left( BG, K(K,3) \right),
\end{equation}
we see $x^* \omega \in {\rm Map}(M, K(K,3)) = H^3(M,K)$,
so the restriction above is that $x^* \omega$ is trivial as an element
of $H^3(M,K)$:
\begin{equation}  \label{eq:app:constr}
x^* \omega \: = \: 1.
\end{equation}
We will examine the implications of this on tori
in section~\ref{app:tori}.

Furthermore, the fiber in ${\rm Map}(M, B \Gamma_{\bullet})$ over
such as $x$ is just the fiber of the Postnikov tower, namely
\begin{equation}
{\rm Map}(M, K(K,2) ) \: = \: H^2(M,K).
\end{equation}
Thus, fibered over every $G$ bundle are the $K$ gerbes, much as one
would expect physically when gauging a 2-group.

\subsection{Computations on tori}
\label{app:tori}

So far we have argued that for both ordinary central extensions and
2-group central extensions of a finite group $G$, for a $G$ bundle to be
in the image of a bundle whose structure group is the extension,
the $G$ bundle must have the property that $x^* \omega = 1$,
where $\omega$ is an element of group cohomology characterizing the
extension, and $x \in {\rm Map}(M,BG)$ encodes the $G$ bundle.

In this section we will unpack that conclusion for the case that $M$ is
a torus.

Recall that for a torus $T$ (of any dimension) we have
\begin{equation}
H^{k}(T,\mathbb{Z}) \: = \: Hom(\wedge^{k}H,\mathbb{Z}),
\end{equation}
where $H = H_{1}(T,\mathbb{Z})$, and the right hand side above can be viewed as skew-symmetric abelian group maps from the direct product of $k$ copies of $H \to \mathbb{Z}$. 

Another way to say this is  as follows.
The cochains  $C^{k}(T,\mathbb{Z})$ are the group of poly linear maps $H^{\times k} \to \mathbb{Z}$, and the cocycles  $Z^{k}(T,\mathbb{Z})$ are the subgroup of maps killed by the Hocschild differential.
Each cocycle is cohomologous to a unique skew-symmetric cocycle and that skew-symmetric cocycle gives a preferred representative in the corresponding cohomology class.

This works up to torsion with arbitrary coefficients. In particular if $K$ is a finite abelian group we have the universal coefficient theorem short exact sequence
\begin{equation}
0 \: \longrightarrow \: {\rm Ext}^{1}(H_{k-1}(T,\mathbb{Z}),K)
\: \longrightarrow \:
 H^{k}(T,K) \: \longrightarrow \: {\rm Hom}(\wedge^{k}H,K) 
\: \longrightarrow \: 0.
\end{equation}

However, note that $H_{k-1}(T,\mathbb{Z})$ is a free finitely generated abelian group and so
\begin{equation}
{\rm Ext}^{1}(H_{k-1}(T,\mathbb{Z}),K)  \: = \: 0,
\end{equation}
which implies
\begin{equation}
H^{k}(T,K) \: = \: {\rm Hom}(\wedge^{k}H,K).
\end{equation}
Thus, for a $K$-valued cocycle, on the torus $T$ its cohomology class is
uniquely determined by its projection  to its skew-symmetric part.

Now, let us consider particular examples.
Earlier in section~\ref{app:class:bundles} we argued that a $G$ bundle
arose from a $\Gamma$ bundle for 
$\Gamma$ an (ordinary) central extension determined by
$\omega \in H^2(G,K)$ if and only if $x^* \omega = 1 \in H^2(M,K)$.
From the analysis above, we see that if $M = T^2$,
$x^* \omega$ is trivial if and only if
\begin{equation}
\frac{ \omega(g_1, g_2) }{ \omega(g_2, g_1) } \: = \: 1
\end{equation}
for commuting pairs $g_1, g_2 \in G$ defining a $G$ bundle (up to
isomorphism).  
In this fashion we recover the constraint
described earlier in section~\ref{sect:ordinary-case}.

For 2-groups and $\omega \in H^3(G,K)$, we can proceed similarly for
$M = T^3$, and see that $x^* \omega = 1$ implies that
\begin{equation} 
\frac{ \omega(g_1, g_2, g_3) }{ \omega(g_1, g_3, g_2) }
\frac{ \omega(g_3, g_1, g_2) }{ \omega(g_3, g_2, g_1) }
\frac{ \omega(g_2, g_3, g_1) }{ \omega(g_2, g_1, g_3) }
\: = \: 1
\end{equation}
as previously outlined in section~\ref{sect:orb-2gp-genl}.

Concretely, given $k$ and $\Gamma$ defined by a central extension
\begin{equation} \label{eq:ext:k}
1 \: \longrightarrow \: B^{k-2} K \longrightarrow \: 
\Gamma \: \longrightarrow \: G \: \longrightarrow \: 1,
\end{equation}
on a torus $T^k$ with $H=H_1(T^k,{\mathbb Z})$,
with central extension corresponding to a class
$[\omega] \in H^{k}(G,K)$,
we have
\begin{equation}
{\rm Hom}(H,\Gamma) \: \longrightarrow \: {\rm Hom}(H,G) \:
\longrightarrow \: {\rm Hom}(\wedge^k H, K),
\end{equation}
where $\wedge^k H \cong {\mathbb Z}$, and so the homomorphism
$\wedge^k H \rightarrow K$ is determined by the element of $K$ which is
the image of $1 \in \wedge^k H = {\mathbb Z}$.
That element of
$K$ is the total skew-symmetrization
of the cocycle $\omega$ when evaluated on the commuting $k$-tuple of
elements of $G$ describing an element of Hom$(H,G)$.  
In other words, for $k=2$, the image of the
generator of $\wedge^2 H$ is
\begin{equation}
\frac{ \omega(g_1, g_2) }{ \omega(g_2,g_1) } ,
\end{equation}
and for $k=3$, the image of the generator of $\wedge^3 H$ is
\begin{equation}
\frac{ \omega(g_1, g_2, g_3) }{ \omega(g_1, g_3, g_2) }
\frac{ \omega(g_3, g_1, g_2) }{ \omega(g_3, g_2, g_1) }
\frac{ \omega(g_2, g_3, g_1) }{ \omega(g_2, g_1, g_3) }.
\end{equation}

We have focused on ordinary groups and 2-groups, but formally
analogous results arise for $k > 3$.  
In particular, for a $k$-torus $T$ and a degree $k$ cohomology class
$[\omega] \in H^k(G,K)$, we will have a similar statement.
From the same analysis as above,
for any $k$, the image of the generator of $\wedge^k H$
\begin{equation}
\prod_{ {\rm permutations} \: \sigma}
\omega\left(g_{\sigma(1)}, g_{\sigma(2)}, \cdots,
g_{\sigma(k)}  \right)^{{\rm sgn} \: \sigma},
\end{equation}
for $[\omega] \in H^k(G,K)$.

In mathematics, discussions of bundles for higher groups and related
notions can be found in e.g.~\cite{wolfson} and references therein.

\section{Decomposition as duality}
\label{app:duality}

In this appendix we describe decomposition, formally,
as a kind of Fourier transform, and then apply that idea to
three-dimensional examples to argue that in orbifolds $[X/\tilde{\Gamma}]$,
where $\tilde{\Gamma}$ is an extension of $G$ by $BK$, the universes
are indexed by representations of $K$ rather than $BK$, which is
what we observe physically.

\subsection{Ordinary decomposition}  \label{app:duality:ordinary}

In this section we will describe decomposition as a form of duality.
This is a special case of the duality discussed in
\cite{dp}; see also \cite{laumon}.

Let $\Gamma$ be a central extension
\begin{equation}  \label{eq:ext1}
1 \: \longrightarrow \: K \: \longrightarrow \: \Gamma \:
\longrightarrow \: G \: \longrightarrow \: 1,
\end{equation}
where both $G$ and $K$ are finite.
Note $[X/\Gamma] \rightarrow [X/G]$ is a $K$ gerbe, a principal
$BK$ bundle.  Now, because $K$ is abelian, $BK = K[1]$ is a stacky
abelian group, and so a principal $BK$ bundle is
a torsor over $BK$, classified by a class in $H^1([X/G],BK) = {\rm Ext}^1_{[X/G]}({\mathbb Z},BK)$.
Hence the extension~(\ref{eq:ext1}) can be interpreted as a complex
of sheaves of abelian groups on $[X/G]$ which is an extension of
${\mathbb Z}$ by $BK$ (over $[X/G]$).

Let ${\cal X}$ be that extension.
It is a family of (complexes of) abelian groups over
$[X/G]$ which sits in an exact sequence
\begin{equation} \label{eq:ses1}
1 \: \longrightarrow \: BK \: \longrightarrow \: {\cal X} \:
\longrightarrow \: {\mathbb Z} \: \longrightarrow \: 1.
\end{equation}
Note that this means ${\cal X}$, as a space (rather than an abelian group)
is a disjoint union of stacks ${\cal X}_n$ over $[X/G]$,
where ${\cal X}_n$ is the preimage of $n \in {\mathbb Z}$.
Each ${\cal X}_n$ is a $K$ gerbe.  For example,
\begin{equation}
{\cal X}_0 \: = \: [X/G] \times BK, \: \: \:
{\cal X}_1 \: = \: [X/\Gamma],
\end{equation}
and for $n > 1$, ${\cal X}_n$ is the $n$th power of ${\cal X}_1 = [X/\Gamma]$
as a $K$ gerbe over $[X/G]$.

Now, we can dualize, by taking homomorphisms into ${\mathbb Z}$.
(Note that the usual dual of $K$ is Hom$(K,{\mathbb Z})$,
whereas the Pontryagin dual of $K$ is Hom$(K, S^1)$.
Also note $S^1 = B {\mathbb Z} = {\mathbb Z}[1]$.)

So, take the short exact sequence~(\ref{eq:ses1}), and dualize to
$B S^1$.  This becomes
\begin{equation}
1 \: \longrightarrow \: {\rm Hom}({\mathbb Z},BS^1) \: \longrightarrow \:
{\rm Hom}({\cal X}, BS^1) \: \longrightarrow \: {\rm Hom}(BK,BS^1)
\: \longrightarrow \: 1.
\end{equation}
(More generally, there are higher Ext's on the right, which can be shown
to vanish here.)

Define the dual group
$\hat{\cal X} = {\rm Hom}({\cal X},BS^1)$, and use the fact that
\begin{eqnarray}
{\rm Hom}({\mathbb Z}, BS^1) & = & BS^1,
\\
{\rm Hom}(BK,BS^1) & = & {\rm Hom}(K,S^1),
\end{eqnarray}
to rewrite the sequence above as
\begin{equation}
1 \: \longrightarrow \: BS^1 \: \longrightarrow \: \hat{\cal X} \:
\longrightarrow \: {\rm Hom}(K,S^1) \: \longrightarrow \: 1.
\end{equation}
Since Hom$(K,S^1)$ is just the characters of $K$, we see that
$\hat{\cal X}$ is a familiy of abelian groups, extending the characters by
$BS^1$, hence is decomposed by characters:
\begin{equation}
\hat{\cal X} \: = \: \coprod_{\lambda} \hat{{\cal X}}_{\lambda},
\end{equation}
where $\hat{\cal X}_{\lambda}$ is an $S^1$-gerbe on $[X/G] \times \lambda$,
for any character $\lambda$.
The part corresponding to ${\cal X}_1$ is $\hat{\cal X}_{\lambda}$ for
$\lambda$ the tautological character.

So far we have discussed ordinary decomposition at a very formal
level as a mathematical duality.  This description has
two ingredients:
\begin{itemize}
\item The data labelling components of the dual, namely characters of
$K$, and
\item the classes of $S^1$ gerbes on $[X/G] \times \lambda$,
which are
\begin{itemize}
\item images under $\lambda$ of the original extension class of $\Gamma$,
\item images under $\lambda$ of the characteristic class of the
principal $BK$ bundle $[X/\Gamma] \rightarrow [X/G]$,
an element of $H^2(G,K)$,
\item images under $\lambda$ of the extension class of ${\mathbb Z}$ by
$BK$, namely Ext$^1({\mathbb Z},BK)$.
\end{itemize}
\end{itemize}

\subsection{Decomposition for two-group extensions}

In this section we will outline a formal understanding of the
decomposition appearing elsewhere in this paper.
In particular, we will argue that, for at least one version of decomposition
for two-group extensions, the universes should be classified by irreducible
representations of $K$, and not\footnote{
See
e.g.~\cite[appendix A]{Schreiber:2008uk}, \cite{gk,Baez:2008hz}
for perspectives on representations of $BK$.
} $BK$.  

Consider the 2-group extension 
\begin{equation}  \label{eq:b2:ext1}
1 \: \longrightarrow \: BK \: \longrightarrow \: \tilde{\Gamma} \:
\longrightarrow \: G \: \longrightarrow \: 1.
\end{equation}
Here, $[X/\tilde{\Gamma}] \rightarrow [X/G]$ is a principal $B^2 K$ bundle.

Note $BK = K[1], B^2 K = K[2]$.

As in section~\ref{app:duality:ordinary}, the extension given by formula~(\ref{eq:b2:ext1})
is equivalent
to specifying a complex of abelian groups $\tilde{\cal X}$ on $[X/G]$
given as
an extension
\begin{equation}  \label{eq:b2:ext2}
1 \: \longrightarrow \: K[2] \: \longrightarrow \:
\tilde{\cal X} \: \longrightarrow \: {\mathbb Z} \: \longrightarrow \: 1.
\end{equation}

Now, we can dualize, but there are several possible targets, such as
${\mathbb Z}$, ${\mathbb Z}[1] = B {\mathbb Z} = S^1$,
${\mathbb Z}[2] = BS^1$, ${\mathbb Z}[3] = B^2 S^1$.

We will `dualize' by taking Hom's into ${\mathbb Z}[3]$.
Applying this to sequence~(\ref{eq:b2:ext2}), we get
\begin{equation}
1 \: \longrightarrow \: {\rm Hom}({\mathbb Z}, {\mathbb Z}[3]) \:
\longrightarrow \: {\rm Hom}(\tilde{\cal X}, {\mathbb Z}[3]) \:
\longrightarrow \: {\rm Hom}(K[2], {\mathbb Z}[3]) \: \longrightarrow
\: 1.
\end{equation}
Define $\widehat{\widetilde{\cal X}} = {\rm Hom}(\tilde{\cal X}, {\mathbb Z}[3])$,
and
note
\begin{eqnarray}
{\rm Hom}( {\mathbb Z}, {\mathbb Z}[3]) & = & {\mathbb Z}[3],
\\
{\rm Hom}( K[2], {\mathbb Z}[3]) & = & {\rm Hom}(K,{\mathbb Z}[1])
\: = \: {\rm Hom}(K,S^1),
\end{eqnarray}
to simplify that sequence to
\begin{equation}
1 \: \longrightarrow \: {\mathbb Z}[3] \: \longrightarrow \:
{\widehat{\widetilde{\cal X}}} \: \longrightarrow \:
{\rm Hom}(K,S^1) \: \longrightarrow \: 1,
\end{equation}
so we see that $\widehat{\widetilde{\cal X}}$ is fibered over characters of
$K$, just as in the previous case.
Note furthermore that
$\widehat{\widetilde{\cal X}}$ is an $S^1$ 2-gerbe over each component,
exactly as expected.

In this section we have made one choice of dualization, dualizing by taking
Hom's to 
${\mathbb Z}[3]$, to understand decomposition.  In principle,
there exist other dualizations, to ${\mathbb Z}[k]$ for other $k$.
We leave an examination of the physical interpretation of such duals,
if any, for future work.

\section{Some results in group cohomology}
\label{app:gpcohom}

In this appendix we collect some results on group cohomology of various
groups, which are used in the main text.

For reference, recall in group cohomology
that coboundaries are determined in degree two by
\begin{equation}
(\delta \alpha)(g_1, g_2, g_3) \: = \:
\frac{ g_1 \cdot \alpha(g_2, g_3) }{ \alpha(g_1 g_2, g_3) }
\frac{ \alpha(g_1, g_2 g_3) }{ \alpha(g_1, g_2) },
\end{equation}
and in degree three by
\begin{equation}
(\delta \omega)(g_1, g_2, g_3, g_4) \: = \:
\frac{ g_1 \cdot \omega(g_2,g_3,g_4) }{ \omega(g_1 g_2, g_3, g_4) }
\frac{ \omega(g_1, g_2 g_3, g_4) }{ \omega(g_1, g_2, g_3 g_4) }
\omega(g_1, g_2, g_3).
\end{equation}

As most of the computations in this paper involve group cohomology with
trivial action on the coefficients, we will assume so unless otherwise
noted.  That said, orientifolds do involve group cohomology with
nontrivial action on the coefficients, so on occasion we will use that
group cohomology instead.

\subsection{${\mathbb Z}_2$}
\label{app:z2}

In this section we will collect some useful results on the group
cohomology of ${\mathbb Z}_2$, which will be useful in setting a pattern
for results later in this appendix for more general products of
${\mathbb Z}_2$'s.

First,
\begin{equation}
H^n( {\mathbb Z}_2, {\mathbb Z}_2) \: = \: {\mathbb Z}_2
\end{equation}
for all (positive) $n$,
where the group cohomology has trivial action on the coefficients, which is
assumed throughout this appendix.  Writing the elements of ${\mathbb Z}_2$
as $\{ 0, 1\}$, the only possibly nonzero normalized cochains are
$x = \omega(1,1,\cdots, 1)$.  In this case,
\begin{equation}
d\omega(1,1,\cdots,1) \: = \:
\left\{ \begin{array}{cl}
2x & n \: {\rm odd}, \\
0 & n \: {\rm even}.
\end{array} \right.
\end{equation}
However, for ${\mathbb Z}_2$ coefficients, $2x=0$, hence $d \omega = 0$ in
all cases.

Note that for $U(1)$ coefficients, for example,
$H^{\rm even}({\mathbb Z}_2, U(1)) = 0$, so the existence of these
cocycles is tied to ${\mathbb Z}_2$ coefficients specifically.

\subsection{${\mathbb Z}_2 \times {\mathbb Z}_2$}
\label{app:z22}

In this section we will collect some useful results on the group cohomology
of ${\mathbb Z}_2 \times {\mathbb Z}_2$.

First, consider the group
\begin{equation}
H^2( {\mathbb Z}_2 \times {\mathbb Z}_2, {\mathbb Z}_2) \: = \:
( {\mathbb Z}_2 )^3.
\end{equation}
We represent the elements as ${\mathbb Z}_2$-valued normalized\footnote{
Throughout this paper, a normalized cocycle is one which is the identity
if any group element among its arguments is the identity.
} cocycles
$C(g,h)$, $g, h \in {\mathbb Z}_2 \times {\mathbb Z}_2$.
Write
\begin{equation}
{\mathbb Z}_2 \times {\mathbb Z}_2 \: = \: 
\{ 1, a, b, ab \},
\end{equation}
and let $x, y, z$ denote the generators of each of the three ${\mathbb Z}_2$'s
in the cohomology group,
then normalized cocycles are listed in table~\ref{table:c:z2z2:z2}.

\begin{table}
\begin{center}
\begin{tabular}{c|cccc}
& $1$ & $a$ & $b$ & $ab$ \\ \hline
$1$ & $1$ & $1$ & $1$ & $1$ \\
$a$ & $1$ & $x$ & $1$ & $x$ \\
$b$ & $1$ & $xyz$ & $y$ & $xz$ \\
$ab$ & $1$ & $yz$ & $y$ & $z$
\end{tabular}
\caption{Representative normalized cocycles for $H^2( {\mathbb Z}_2 \times
{\mathbb Z}_2, {\mathbb Z}_2)$.
For example, $C(a,b) = 1$, $C(b,a) = xyz$.
\label{table:c:z2z2:z2}}
\end{center}
\end{table}

In particular, for ${\mathbb Z}_2$ coefficients and normalized cocycles,
$C(g,g)$ is coboundary-invariant, and from
table~\ref{table:c:z2z2:z2}, we see that
\begin{equation}
C(a,a) \: = \: x, \: \: \:
C(b,b) \: = \: y, \: \: \:
C(ab,ab) \: = \: z
\end{equation}
naturally encode the generators of each of the three ${\mathbb Z}_2$'s
in $H^2( {\mathbb Z}_2 \times {\mathbb Z}_2, {\mathbb Z}_2)$.

The cohomology groups of ${\mathbb Z}_2 \times {\mathbb Z}_2$ also include
\begin{equation}
H^2( ({\mathbb Z}_2)^2 , U(1)) \: = \: {\mathbb Z}_2,
\: \: \:
H^3( ({\mathbb Z}_2)^2 , U(1)) \: = \: ( {\mathbb Z}_2)^3,
\: \: \:
H^4( ({\mathbb Z}_2)^2 , U(1)) \: = \:
{\mathbb Z}_2 \times {\mathbb Z}_2.
\end{equation}
Degree three cohomology was recently discussed in detail
in \cite{Robbins:2021lry} and \cite[appendix A]{Robbins:2021xce},
giving both representatives as well as invariants
that distinguish different cohomology classes.

For use elsewhere, let us characterize the elements of $H^4(
{\mathbb Z}_2 \times {\mathbb Z}_2, U(1))$ more precisely.
Let $\alpha$ denote a normalized 4-cocycle, meaning
$\alpha(g_1, g_2, g_3, g_4) = 1$ if any $g_i = 1$.
(In effect, this is a gauge choice, which requires in
evaluating coboundaries that 3-cochains equal $1$ if any of their
arguments is $1$.)

If we write each $g \in {\mathbb Z}_2 \times {\mathbb Z}_2$ as
$g = (x,y)$ for $x,y \in \{0,1\}$, then normalized cocycles
$\alpha_{0,\cdots, 3}$
representing different elements of $H^4( {\mathbb Z}_2 \times {\mathbb Z}_2,
U(1))$ are as follows:
\begin{eqnarray}
\alpha_0( (x_1, y_1), (x_2, y_2), (x_3, y_3), (x_4, y_4) )
& = & +1,
\\
\alpha_1( (x_1, y_1), (x_2, y_2), (x_3, y_3), (x_4, y_4) )
& = &
(-1)^{x_1 y_2 y_3 y_4},
\\
\alpha_2( (x_1, y_1), (x_2, y_2), (x_3, y_3), (x_4, y_4) )
& = &
(-1)^{x_1 x_2 x_3 y_4},
\\
\alpha_3( (x_1, y_1), (x_2, y_2), (x_3, y_3), (x_4, y_4) )
& = &
(-1)^{x_1 y_4 (x_2 x_3 + y_2 y_3)}.
\end{eqnarray}
As elements of ${\mathbb Z}_2 \times {\mathbb Z}_2$,
$\alpha_0$ is the identity and
$\alpha_3 = \alpha_1 \alpha_2$.

For any pair $(g,h) \in {\mathbb Z}_2 \times {\mathbb Z}_2$,
we can define an invariant $A(g,h)$ of normalized 4-cocycles,
invariant under coboundaries, as
\begin{equation} \label{eq:z2z2:invt}
A(g,h) \: = \:
\frac{ \alpha(g,g,g,h) }{ \alpha(g,g,h,g) }
\frac{ \alpha(g,h,g,g) }{ \alpha(h,g,g,g) }.
\end{equation}
Applying these invariants to the normalized cocycles above,
and writing
${\mathbb Z}_2 \times {\mathbb Z}_2 = \{1, a, b, ab\}$,
with $a = (1,0)$, $b = (0,1)$,
we compute invariants corresponding to elements of $H^4({\mathbb Z}_2 \times
{\mathbb Z}_2, U(1))$ as in table~\ref{table:h4:z2z2:invt}.
This can be useful in distinguishing elements of $H^4({\mathbb Z}_2 \times
{\mathbb Z}_2,U(1))$, as cocycles are only defined up to coboundaries.

\begin{table}
\begin{center}
\begin{tabular}{c|cccccc}
$\alpha$ & $A(a,b)$ & $A(a,ab)$ & $A(b,a)$ & $A(b,ab)$ & $A(ab,a)$ & $A(ab,b)$
\\ \hline
$\alpha_0$ & $+1$ & $+1$ & $+1$ & $+1$ & $+1$ & $+1$ \\
$\alpha_1$ & $+1$ & $+1$ & $-1$ & $-1$ & $-1$ & $-1$
\\
$\alpha_2$ & $-1$ & $-1$ & $+1$ & $+1$ & $-1$ & $-1$
\\
$\alpha_3$ & $-1$ & $-1$ & $-1$ & $-1$ & $+1$ & $+1$
\end{tabular}
\caption{Invariants computed from elements of $H^4({\mathbb Z}_2 \times
{\mathbb Z}_2, U(1))$.
\label{table:h4:z2z2:invt} }
\end{center}
\end{table}

\subsection{${\mathbb Z}_2 \times {\mathbb Z}_2 \times {\mathbb Z}_2$}
\label{app:z23}

Write
$\Gamma = ( {\mathbb Z}_2)^3 = \langle a, b, c \rangle$,
$G = {\mathbb Z}_2 = \langle a \rangle$,
$K = ({\mathbb Z}_2)^2 = \langle b, c \rangle$.

Now,
\begin{equation}
H^3( \Gamma, U(1) ) \: = \: ( {\mathbb Z}_2 )^7,
\end{equation}
which we can understand as arising from
the Lyndon-Hochschild-Serre spectral sequence as\footnote{
Since $\Gamma$ is just a direct sum, the extension class vanishes,
and so all of the maps $d_n$ in the spectral sequence are trivial and
so the sequence stabilizes at $E_2^{p,1}$.
}
\begin{eqnarray}
H^3(K, U(1)) & = & ( {\mathbb Z}_2)^3,
\\
H^1(G, H^2(K,U(1))) & = & H^1( {\mathbb Z}_2, {\mathbb Z}_2) \: = \:
{\mathbb Z}_2,
\\
H^2(G, H^1(K,U(1))) & = & H^2( {\mathbb Z}_2, ( {\mathbb Z}_2)^2 )
\: = \:
( {\mathbb Z}_2)^2,
\\
H^3(G, U(1)) & = & {\mathbb Z}_2.
\end{eqnarray}

Thus three of the generators of $H^3(\Gamma,U(1))$, call them $\omega_1$, $\omega_2$, $\omega_3$,
are pullbacks from $H^3(K, U(1))$ under the projection
$\Gamma = G \times K \rightarrow K$.

Another generator, call it $\omega_7$, is similarly
a pullback from $H^3(G,U(1))$, and is given by
\begin{equation}
\omega_7((a_1,b_1,c_1), (a_2,b_2,c_2), (a_3,b_3,c_3)) = (-1)^{a_1 a_2 a_3}
\end{equation}
where here we identify $a, b,c \in \{ 0, 1 \}$.

One more generator, call it $\omega_4$, comes from $H^1(G, H^2(K,U(1)))$ and can be
represented as
\begin{equation}
\omega_4 = (-1)^{a_1 b_2 c_3}.
\end{equation}

The final two generators of $H^3(\Gamma,U(1))$, call them $\omega_{5,6}$, come from $H^2(G, H^1(K,U(1)))$, and
can be represented as
\begin{eqnarray}
\omega_5 & = & (-1)^{a_1 a_2 b_3},
\\
\omega_6 & = & (-1)^{a_1 a_2 c_3}.
\end{eqnarray}

One can check that all of the $\omega_i$'s are co-closed
and that they are not cohomologous
(they differ on coboundary invariants $\omega(g,g,g)$
for some $g \in \Gamma$).
Furthermore, it is also easy to check that the
$C$ field discrete torsion phase~(\ref{eq:dw-phases}) on $T^3$
are nontrivial for $\omega_4$ evaluated on triples of the form
$(a_1x, b_2y, c_3z)\}$ and their permutations for 
\begin{equation}
x \in \{1, b_1, c_1, b_1 c_1\}, \: \: \:
y \in \{1, a_2, c_2, a_2 c_2 \}, \: \: \:
z \in \{1, a_2, b_2, a_2 b_2 \}.
\end{equation}

The group
\begin{equation}
H^3( ({\mathbb Z}_2)^3, {\mathbb Z}_2)
\: = \: ( {\mathbb Z}_2 )^{10}.
\end{equation}
From Lyndon-Hochschild-Serre as before, we can write this as
\begin{eqnarray}
H^3( K, {\mathbb Z}_2 ) & = &
H^3( ( {\mathbb Z}_2)^2, {\mathbb Z}_2) \: = \:
( {\mathbb Z}_2)^4,
\\
H^1(G, H^2(K, {\mathbb Z}_2)) & = &
{\rm Hom}( {\mathbb Z}_2, ( {\mathbb Z}_2)^3 ) \: = \:
( {\mathbb Z}_2)^3,
\\
H^2(G, H^1(K, {\mathbb Z}_2)) & = &
H^2( {\mathbb Z}_2, ( {\mathbb Z}_2)^2 ) \: = \:
( {\mathbb Z}_2 )^2,
\\
H^3(G, {\mathbb Z}_2) & = & {\mathbb Z}_2.
\end{eqnarray}

\subsection{$({\mathbb Z}_2)^k$}
\label{app:genl}

In this appendix we give a basis of cocycles for
$H^n(({\mathbb Z}_2)^k, {\mathbb Z}_2)$ for any $n$, $k$.

Represent $g \in ( {\mathbb Z}_2)^k$ as
$g = (x^1, \cdots, x^k)$ with $x^i \in \{0, 1\}$.  Pick $k$ nonnegative
integers $m_1, \cdots, m_k$ such that
\begin{equation}
m_1 \: + \: m_2 \: + \: \cdots \: + \: m_k \: = \: n.
\end{equation}
There will be
\begin{equation}
N \: = \: \left( \begin{array}{c} n+k-1 \\ k-1 \end{array} \right)
\end{equation}
possibilities, each of which corresponds to
a cocycle.  In particular, we will see that
\begin{equation}
H^n( ({\mathbb Z}_2)^k, {\mathbb Z}_2) \: = \: ( {\mathbb Z}_2)^N.
\end{equation}
Define a function
\begin{equation}
f_{m}: \: \{1, \cdots, n \} \: \longrightarrow \: \{1, \cdots, k\}
\end{equation}
($m \in \{1, \cdots, N\}$) by
\begin{equation}
f_m(a) \: = \: j
\end{equation}
for $j$ such that
\begin{equation}
m_1 + m_2 + \cdots +
m_{j-1} < a \leq m_1 + m_2 + \cdots + m_j 
\end{equation}
(in conventions in which
$m_0 = 0$).

Then define
\begin{equation}
\omega_{m}(g_1, \cdots, g_n) \: = \: (-)^{\alpha}
\end{equation}
for
\begin{equation}
\alpha \: = \:
\prod_{a=1}^n x_a^{f_m(a)} .
\end{equation}

For example, consider $H^{n}({\mathbb Z}_2,{\mathbb Z}_2)$.
In this case, $N=1$ for all $n$, and $f_1(a) = 1$ for all $a \in
\{1, \cdots, n\}$.  In each case, if we write
${\mathbb Z}_2 = \langle a \rangle$, then
a normalized cocycle for $H^n( {\mathbb Z}_2, {\mathbb Z}_2)$ is
$(-)^a$, for any $n$.

For another example, consider the group cohomology
of $({\mathbb Z}_2)^2 = \langle a, b \rangle$,
starting with $H^2( ({\mathbb Z}_2)^2, {\mathbb Z}_2)$.
Here, $N=3$, corresponding to the three sums
\begin{equation}
1+1, \: \: \: 2+0, \: \: \: 0+2.
\end{equation}
Corresponding respectively to those three sums we have the functions
\begin{equation}
\begin{array}{c}
f_{1+1}(1) \: = \: 1, \: \: \:
f_{1+1}(2) \: = \: 2,
\\
f_{2+0}(1) \: = \: 1, \: \: \:
f_{2+0}(2) \: = \: 1,
\\
f_{0+2}(1) \: = \: 2, \: \: \:
f_{0+2}(2) \: = \: 2,
\end{array}
\end{equation}
which correspond to the three cocycles
\begin{equation}
(-)^{a_1 b_2}, \: \: \: (-)^{a_1 a_2}, \: \: \:
(-)^{b_1 b_2}.
\end{equation}

One can compute $H^3( ({\mathbb Z}_2)^2, {\mathbb Z}_2)$ similarly.
Here, $N=4$, corresponding to the four sums
\begin{equation}
2+1, \: \: \:
1+2, \: \: \:
3+0, \: \: \:
0+3,
\end{equation}
and corresponding to those sums are the functions
\begin{equation}
\begin{array}{c}
f_{2+1}(1) \: = \: 1, \: \: \:
f_{2+1}(2) \: = \: 1, \: \: \:
f_{2+1}(3) \: = \: 2,
\\
f_{1+2}(1) \: = \: 1, \: \: \:
f_{1+2}(2) \: = \: 2, \: \: \:
f_{1+2}(3) \: = \: 2,
\\
f_{3+0}(1) \: = \: 1, \: \: \:
f_{3+0}(2) \: = \: 1, \: \: \:
f_{3+0}(3) \: = \: 1,
\\
f_{0+3}(1) \: = \: 2, \: \: \:
f_{0+3}(2) \: = \: 2, \: \: \:
f_{0+3}(3) \: = \: 2.
\end{array}
\end{equation}
The corresponding cocycles are
\begin{equation}
\omega_{2+1} \: = \: (-)^{a_1 a_2 b_3}, \: \: \:
\omega_{1+2} \: = \: (-)^{a_1 b_2 b_3}, \: \: \:
\omega_{3+0} \: = \: (-)^{a_1 a_2 a_3}, \: \: \:
\omega_{0+3} \: = \: (-)^{b_1 b_2 b_3}.
\end{equation}

For another example, for $({\mathbb Z}_2)^3 = \langle a, b, c \rangle$,
there is a basis of cocycles given by
\begin{eqnarray}
H^1( ({\mathbb Z}_2)^3, {\mathbb Z}_2)
& = &
\{ (-)^a, (-)^b, (-)^c \} \mbox{  (3 elements)},
\\
H^2( ({\mathbb Z}_2)^3, {\mathbb Z}_2)
& = &
\{  (-1)^{a_1 a_2}, (-1)^{a_1 b_2}, (-1)^{a_1 c_2}, (-1)^{b_1 b_2}, (-1)^{b_1 c_2}, (-1)^{c_1 c_2} \}
\nonumber \\
& & \hspace*{0.5in}
\mbox{ (6 elements)},
\\
H^3( ({\mathbb Z}_2)^3, {\mathbb Z}_2)
& = &
\{
(-1)^{a_1 a_2 a_3}, (-1)^{a_1 a_2 b_3}, (-1)^{a_1 a_2 c_3}, 
(-1)^{a_1 b_2 b_3}, (-1)^{a_1 b_2 c_3}, (-1)^{a_1 c_2 c_3}, 
\nonumber \\
& & \hspace*{0.5in}
(-1)^{b_1 b_2 b_3}, (-1)^{b_1 b_2 c_3}, (-1)^{b_1 c_2 c_3}, (-1)^{c_1 c_2 c_3}
\}
\nonumber \\
& & \hspace*{0.5in}
 \mbox{ (10 elements)}.
\end{eqnarray}
Furthermore, $H^n( ({\mathbb Z}_2)^3, {\mathbb Z}_2)$ has
a basis of
\begin{equation}
\left( \begin{array}{c} n+2 \\ 2 \end{array} \right) \: = \: 
\frac{(n+1)(n+2)}{2}
\end{equation}
cocycles.

\subsection{Dihedral groups}

Let $D_n$ denote the dihedral group of order $2n$.
Then,
\begin{equation}
H^3(D_n,U(1)) \: = \: \left\{ \begin{array}{cl}
{\mathbb Z}_{2n} & $n$ \mbox{ odd}, \\
({\mathbb Z}_2)^2 \times {\mathbb Z}_n & $n$ \mbox{ even}.
\end{array} \right.
\end{equation}

Below we give an explicit cocycle for $H^2(D_4,Z_2) = ( {\mathbb Z}_2)^3$.
The group $D_4$ is presented in terms of generators $a$, $b$, with relations
\begin{equation}
a^2 \: = \: 1 \: = \: b^4, \: \: \:
a b a = b^{-1} = b^3,
\end{equation}
and the results are expressed in terms of $x, y, z$ which generate each of
three ${\mathbb Z}_2$'s in table~\ref{table:h2-d4-z2}.

\begin{table}[h]
\begin{center}
\begin{tabular}{c|cccccccc}
 & $1$ & $b$ & $b^2$ & $b^3$ & $a$ & $ba$ & $b^2a$ & $b^3 a$  \\ \hline
$1$ & $1$ & $1$ & $1$ & $1$ & $1$ & $1$ & $1$ & $1$ \\
$b$ & $1$ & $1$ & $1$ & $x$ & $1$ & $1$ & $1$ & $x$ \\
$b^2$ & $1$ & $1$ & $x$ & $x$ & $1$ & $1$ & $x$ & $x$ \\
$b^3$ & $1$ & $x$ & $x$ & $x$ & $1$ & $x$ & $x$ & $x$ \\
$a$ & $1$ & $y$ & $x$ & $y$ & $z$ & $yz$ & $xz$ & $yz$ \\
$ba$ & $1$ & $xy$ & $x$ & $y$ & $z$ & $xyz$ & $xz$ & $yz$ \\
$b^2a$ & $1$ & $xy$ & $1$ & $y$ & $z$ & $xyz$ & $z$ & $yz$ \\
$b^3a$ & $1$ & $xy$ & $1$ & $xy$ & $z$ & $xyz$ & $z$ & $xyz$
\end{tabular}
\caption{Table of cocycles representing elements of $H^2(D_4,{\mathbb Z}_2)
= ( {\mathbb Z}_2)^3$.  The variables $x$, $y$, $z$ generate the three
${\mathbb Z}_2$'s.
\label{table:h2-d4-z2}
}
\end{center}
\end{table}

\begin{table}[h]
\begin{center}
\begin{tabular}{c|cccccccc}
 & $1$ & $b$ & $b^2$ & $b^3$ & $a$ & $ba$ & $b^2a$ & $b^3 a$  \\ \hline
$1$ & $1$ & $1$ & $1$ & $1$ & $1$ & $1$ & $1$ & $1$ \\
$b$ & $1$ & $1$ & $1$ & $1$ & $-$ & $-$ & $-$ & $-$ \\
$b^2$ & $1$ & $1$ & $1$ & $1$ & $x$ & $x$ & $x$ & $x$ \\
$b^3$ & $1$ & $1$ & $1$ & $1$ & $-$ & $-$ & $-$ & $-$ \\
$a$ & $1$ & $-$ & $x$ & $-$ & $1$ & $-$ & $x$ & $-$ \\
$ba$ & $1$ & $-$ & $x$ & $-$ & $-$ & $1$ & $-$ & $x$ \\
$b^2a$ & $1$ & $-$ & $x$ & $-$ & $x$ & $-$ & $1$ & $-$ \\
$b^3a$ & $1$ & $-$ & $x$ & $-$ & $-$ & $x$ & $-$ & $1$
\end{tabular}
\caption{Table of invariant phases $\epsilon(g,h) = C(g,h)/C(h,g)$ using the
cocycles in table~\ref{table:h2-d4-z2}.
An entry `$-$' indicates a non-commuting pair.
\label{table:phases:d4-z2}
}
\end{center}
\end{table}

\end{document}